\documentclass[10pt,journal,compsoc]{IEEEtran}
\ifCLASSOPTIONcompsoc
  \usepackage[nocompress]{cite}
\else
  \usepackage{cite}
\fi
\ifCLASSINFOpdf
\else
\fi
\usepackage{amsmath}
\DeclareMathOperator*{\argmax}{arg\,max}
\usepackage{etoolbox,xspace}

\usepackage{framed}
\usepackage[shortlabels]{enumitem}
\usepackage{amsmath}
\usepackage[ruled,vlined,linesnumbered]{algorithm2e}
\usepackage{multirow}
\usepackage{bigstrut}
\usepackage{amssymb}
\usepackage{amsthm}
\usepackage{booktabs}
\usepackage{graphicx}
\usepackage{subfigure}

\usepackage{colortbl}
\usepackage[figuresright]{rotating}

\usepackage{listings}
\usepackage[ruled,vlined,linesnumbered]{algorithm2e}
\SetCommentSty{mycommfont}

\usepackage{multirow}
\usepackage{booktabs}
\usepackage{comment}
\usepackage[T1]{fontenc} 
\usepackage{array}
\usepackage{url}
\usepackage{pifont}
\usepackage{latexsym}
\usepackage{flushend}
\usepackage{ragged2e}

\usepackage{lipsum}

\definecolor{mygreen}{rgb}{0,0.5,0}
\definecolor{myblue}{rgb}{0,0,1}
\definecolor{mymauve}{rgb}{0.58,0,0.82}
\definecolor{myblack}{rgb}{0.24,0.17,0.12}
\definecolor{awesome}{rgb}{1.0, 0.13, 0.32}
\usepackage{xcolor}
\usepackage{ragged2e}

\usepackage[normalem]{ulem} 
\newcommand\hl{\bgroup\markoverwith
	{\textcolor{lightgray}{\rule[-.5ex]{2pt}{2.5ex}}}\ULon}

 \usepackage[hidelinks]{hyperref}

\definecolor{light-gray}{gray}{0.80}
\usepackage{varwidth}
\usepackage{threeparttable}
\usepackage[utf8]{inputenc}
\usepackage[most]{tcolorbox}
\usepackage{booktabs}

\definecolor{light-gray}{gray}{0.80}
\definecolor{light-green}{RGB}{254,242,185 }
\definecolor{light-blue}{RGB}{252,207,167}

\def\BibTeX{{\rm B\kern-.05em{\sc i\kern-.025em b}\kern-.08em
    T\kern-.1667em\lower.7ex\hbox{E}\kern-.125emX}}

\newcommand*\circled[1]{\tikz[baseline=(char.base)]{
		\node[shape=circle,draw,inner sep=0.1pt,fill=black, text=white] (char) {#1};}}
\newcommand{\ourSol}{\textsc{LLM4CBI}\xspace}

\newcommand{\existPrompt}{\textsc{LLM4CBI}$_{ep}$\xspace}
\newcommand{\specficPrompt}{\textsc{LLM4CBI}$_{sp}$\xspace}
\newcommand{\randSel}{\textsc{LLM4CBI}$_{rand}$\xspace}
\newcommand{\selNoVal}{\textsc{LLM4CBI}$_{selnov}$\xspace}

\newcommand{\gptFourLLMs}{\textsc{LLM4CBI}(GPT-4)\xspace}

\newcommand{\alpLLMs}{\textsc{LLM4CBI}(Alpaca)\xspace}
\newcommand{\vicLLMs}{\textsc{LLM4CBI}(Vicuna)\xspace}

\newcommand{\allLLMs}{\textsc{LLM4CBI}(GPT4ALL)\xspace}

\usepackage{threeparttable}
\usepackage{balance}

\begin{document}

%
\title{Isolating Compiler Bugs by Generating Effective Witness Programs with Large Language Models}
%
%
%
%

\author{
        Haoxin Tu,
        Zhide Zhou,
        He Jiang$^*$,\thanks{* He Jiang is the corresponding author.} 
        Imam Nur Bani Yusuf, 
        Yuxian Li,
        Lingxiao Jiang
        
\IEEEcompsocitemizethanks{
\IEEEcompsocthanksitem H. Tu is with the School of Software, Dalian University of Technology, Dalian, China. H. Tu is also with the School of Computing and Information Systems, Singapore Management University, Singapore. E-mail: haoxintu@gmail.com.
\IEEEcompsocthanksitem Z. Zhou and H. Jiang are with the School of Software, Dalian University of Technology, Dalian, China, and Key Laboratory for Ubiquitous Network and Service Software of Liaoning Province. H. Jiang is also with DUT Artificial Intelligence, Dalian, China. E-mail: jianghe@dlut.edu.cn, cszide@gmail.com.
\IEEEcompsocthanksitem I. Yusuf, Y. Li, and L. Jiang are with the School of Computing and Information Systems, Singapore Management University, Singapore. E-mails: imamy.2020@phdcs.smu.edu.sg, liyuxianjnu@gmail.com, lxjiang@smu.edu.sg.

}
}

%
%

\markboth{IEEE TRANSACTIONS ON SOFTWARE ENGINEERING}%
{Shell \MakeLowercase{\textit{et al.}}: Bare Demo of IEEEtran.cls for Computer Society Journals}
%



\IEEEtitleabstractindextext{%
\begin{abstract}
\justifying
Compiler bugs pose a significant threat to safety-critical applications, and promptly as well as effectively isolating these bugs is crucial for assuring the quality of compilers. However, the limited availability of debugging information on reported bugs complicates the compiler bug isolation task. Existing compiler bug isolation approaches convert the problem into a test program mutation problem, but they are still limited by ineffective mutation strategies or high human effort requirements.
Drawing inspiration from the recent progress of pre-trained Large Language Models (LLMs), such as ChatGPT, in code generation, we propose a new approach named \ourSol to utilize LLMs to generate effective test programs for compiler bug isolation. However, using LLMs directly for test program mutation may not yield the desired results due to the challenges associated with formulating precise prompts and selecting specialized prompts. To overcome the challenges, three new components are designed in \ourSol.
First, \ourSol utilizes a program complexity-guided prompt production component, which leverages data and control flow analysis to identify the most valuable variables and locations in programs for mutation. Second, \ourSol employs a memorized prompt selection component, which adopts reinforcement learning to select specialized prompts for mutating test programs continuously. 
Third, a test program validation component is proposed to select specialized feedback prompts to avoid repeating the same mistakes during the mutation process. 
Compared with the state-of-the-art approaches (DiWi and RecBi) over 120 real bugs from the two most popular compilers, namely GCC and LLVM, our evaluation demonstrates the advantages of \ourSol:
It can isolate 69.70\%/21.74\% and 24.44\%/8.92\% more bugs than DiWi and RecBi within Top-1/Top-5 ranked results.
Additionally, we demonstrate that the LLMs component (i.e., GPT-3.5) used in \ourSol can be easily replaced by other LLMs while still achieving reasonable results in comparison to related studies.
\end{abstract}

}

\maketitle

\IEEEdisplaynontitleabstractindextext

%
\maketitle
\section{Introduction} \label{sec:introduction}
\IEEEPARstart{C}{ompilers} serve a fundamental role in 
building reliable software systems, and bugs in compilers can have catastrophic consequences \cite{chen2020survey,csmith}. To mitigate such threats, a crucial task is to isolate the bugs promptly and effectively. 
However, isolating compiler bugs poses significant challenges due to the limited debugging information.
A prevalent direction for resolving the problem is to transform the bug isolation problem into a test program mutation problem \cite{diwi,recbi}. The core idea behind such approaches is first to generate a set of witness (i.e., passing) test program mutants by mutating the given failing test program and then collecting the passing and failing spectrum. Finally, combined with Spectrum-based Fault Localization (SBFL) techniques, suspicious files are ranked. 

However, existing program mutation strategies in DiWi \cite{diwi} and RecBi \cite{recbi} exhibit limitations in terms of effectiveness and demand substantial human effort. First, these strategies are limited in generating diverse test programs. DiWi only supports the local mutation, such as changing the type of a variable. Although RecBi supports more mutation strategies, such as structural mutation, i.e., changing the control flow of the program, it can only synthesize statement conditions ingredients without statement body (see more details in Section \ref{sec:motivating-example}). Moreover, the random selection of variables and locations for mutation in both DiWi and RecBi is limited by the diversity of the generated test programs. 
Second, the mutation process in DiWi and RecBi requires significant human effort. Before mutation, DiWi and RecBi require manual code additions to collect necessary context information, and RecBi necessitates the construction of ingredients from existing test programs. Additionally, these strategies pay insufficient attention to the validity of the generated programs that contain undefined behaviors, thereby reducing their effectiveness of bug isolation. Overall, these limitations in ineffective mutation and the associated human effort emphasize the need for improved program mutation strategies in bug isolation.

Inspired by the recent progress of pre-trained Large Language Models (LLMs) (e.g., ChatGPT \cite{chatgpt}) in code generation, we propose \ourSol, i.e., pre-trained \textbf{L}arge \textbf{L}anguage \textbf{M}odels for \textbf{C}ompiler \textbf{B}ug \textbf{I}solation. Our key insights are that (1)
LLMs are trained with large-scale datasets of code, so the test programs produced by LLMs tend to be diverse; (2) LLMs have a good learning \& reflection mechanism to help generate better outputs based on the users' feedback following a prompt-response dialog paradigm; (3) prompts used by LLMs can be expressed in a natural language, easier for users to use and reducing the human effort to finish a task. Thus, adapting LLMs to generate effective test programs could be promising. 
However, directly using LLMs to generate effective test programs for compiler bug isolation presents several difficulties, raising the following two challenges that need to be addressed.

\smallskip
\noindent
\textbf{\textit{Challenge 1: Formulation of Precise Prompts}.} The quality of prompts plays a crucial role in employing the program mutation capabilities of LLMs \cite{liu2022fill}, but using existing natural mutation descriptions as prompts may not be effective. For example, the mutation rule ``\textit{insert an if statement}'' is a mutation description used in existing work RecBi \cite{recbi}. Such description lacks precision on {\it which} variables to use and {\it where} to insert the statement, limiting the possibility of mutating failing test programs (i.e., those that can trigger the bug) into passing ones (i.e., those that can not trigger the bug).
Due to the difficulties in selecting precise variables and locations, the existing approaches DiWi \cite{diwi} and RecBi \cite{recbi} randomly select them, which is shown to be ineffective (see more evaluation results in Section \ref{sec:answer2}). Therefore, it is necessary to formulate precise mutation prompts to assist LLMs in generating effective test programs.

\smallskip
\noindent
\textbf{\textit{Challenge 2: Selection of Specialized Prompts}.} When several prompts are collected, selecting the specialized ones for mutating specific failing test programs is important and challenging. The reasons are two-fold. First, compiler bugs tend to be different and have different language features. One prompt may be useful for mutating one particular failing test program but may not be helpful for another failing test program. Randomly selecting prompts to mutate test programs may be ineffective (see more evaluation results in Section \ref{sec:answer2}). Second, LLMs may make different mistakes when mutating different test programs, and different feedback prompts should also be given to different test programs. Therefore, a new prompt selection strategy is needed to select specialized prompts.

To overcome the above challenges, three new components are designed in \ourSol to utilize LLMs for generating effective test programs for compiler bug isolation. First, a {\bf precise prompt production component} is designed to address the first challenge. A precise prompt pattern that could accurately represent the desired mutation operations is first introduced. Then, program complexity metrics measured by the data and control flow analysis are utilized to identify the most relevant variables and optimal insertion locations.
Second, two new components, i.e., a {\bf memorized prompt selection component} and a {\bf lightweight test program validation component}, are proposed for selecting specialized prompts. In the prompt selection component, \ourSol incorporates memorized search via reinforcement learning to track and accumulate rewards based on the performance of LLMs, which allows \ourSol to continually select the specialized prompts for mutating specific test programs. In the test programs validation component, \ourSol leverages a static analysis to detect and filter out potential invalid test programs that may contain undefined behaviors, thus mitigating the risks associated with invalid programs.

Empirical evaluations over 120 real bugs from the two most popular C open-source compilers, i.e., GCC and LLVM, 
are conducted
to demonstrate the effectiveness of \ourSol. First, we have compared \ourSol with two state-of-the-art approaches (i.e., DiWi \cite{diwi} and RecBi \cite{recbi}) in terms of compiler bug isolation capabilities. The results show \ourSol can isolate 69.70\%/21.74\% and 24.44\%/8.92\% more bugs than DiWi and RecBi within Top-1/Top-5 ranked results, respectively.
Second, we have evaluated the effectiveness of three new components of \ourSol, and the results show that all the components contribute to the effectiveness of \ourSol.
Third, we demonstrate that \ourSol is extensible, i.e., the LLMs component (i.e., GPT-3.5) used in \ourSol can be easily replaced by other LLMs (e.g., Alpaca \cite{alpaca}, Vicuna \cite{vicuna}, and GPT4ALL \cite{gpt4all}) while still achieving reasonable results in comparison to related studies.

\smallskip
\noindent \textbf{Contributions.} We make the following contributions:
\begin{itemize}[leftmargin=1em,nosep]
    \item To our knowledge, \ourSol is the first work aiming to leverage the capabilities of LLMs for compiler bug isolation tasks in the field.
    \item Three new components, i.e., precise prompt production, memorized prompt selection, and lightweight test program validation, are proposed to guide LLMs to generate effective test programs for compiler bug isolation.
    \item Our empirical evaluation demonstrates that \ourSol is not only effective for compiler bug isolation but also extensible because other LLMs can be easily integrated into \ourSol.
    \item  \ourSol\footnote{The source code of \ourSol is publicly available at \url{https://github.com/haoxintu/LLM4CBI}.} paves the way for future research in compiler bug isolation, opening exciting opportunities to further explore and leverage the capabilities of LLMs for more efficient and effective bug isolation techniques.
\end{itemize}

\smallskip
\noindent \textbf{Organizations.}
Section \ref{sec:motivation} gives the background and our motivation. 
Section \ref{sec:approach} describes
the design of \ourSol.
Section \ref{sec:evaluation} presents the implementation and the evaluation results.
Section \ref{sec:discussion} and \ref{sec:threats} discuss the limitations of our approach and threats to validity.
Section \ref{sec:related-work} describes related work, and
Section \ref{sec:conclusion} concludes with future work.

\section{Background and Motivation} \label{sec:motivation}

In this section, we first give the background about test program mutation for compiler bug isolation and Large Language Models (LLMs). Then, we use an example to 
illustrate the limitations of existing approaches and highlight the advantages of our approach.

\subsection{Background}

\subsubsection{Test Program Mutation for Compiler Bug Isolation} \label{sec:cbi}
\begin{figure}[t]
\centering
\includegraphics[width=0.8\linewidth]{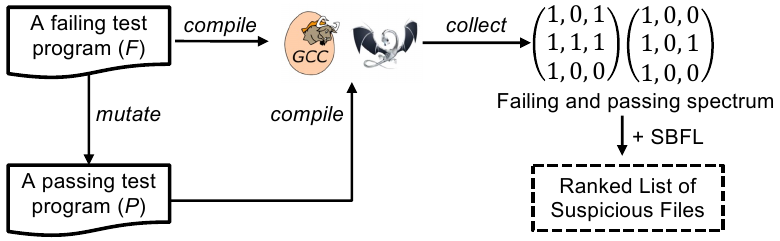}
\vspace{-1em}
\caption{General workflow of existing compiler bug isolation}
\label{fig:cbi}
\end{figure}

Bug isolation is one of the most important activities in software debugging \cite{zhang2024contextaug,xuan2014test,zhang2024improving,renieres2003fault,moon2014ask}.
Fig. \ref{fig:cbi} exemplifies the prevalent workflow of existing compiler bug isolation approaches \cite{diwi,recbi}. Given a failing test program that can trigger the compiler bug, existing approaches first use different mutation strategies to produce a passing test program that does not trigger any bugs (we refer to such passing test programs to witness test programs following existing studies \cite{diwi,recbi}).
Then, both the failing and passing test programs are subjected to compilation, enabling the collection of code coverage information of the compiler source files.
All the compiler files that are covered by a failing test program during compilation are considered suspicious. Conversely, the passing test program serves to mitigate suspicions regarding innocent files that may have been implicated. To eventually isolate the buggy files, following the well-established principles of Spectrum-Based Fault Localization (SBFL) \cite{abreu2007accuracy,wong2016survey}, 
they compare the execution traces (or spectra) between failing test programs and passing test programs using a formula such as Ochiai \cite{abreu2007accuracy}. Two recent studies, i.e., DiWi \cite{diwi} and RecBi \cite{recbi}, follow the same strategy, and their goal is to \textit{generate a set of programs that have slightly different control- and data-flow information compared with the falling test program to flip the compiler execution results (i.e., from failing to passing).} We aim to achieve the same goal by leveraging a new approach based on LLMs in this study.

\subsubsection{Suspicious Files Ranking}

\ourSol utilizes the concept of SBFL (Spectrum-Based Fault Localization) to identify potentially buggy compiler files by comparing the coverage of failing and passing tests, as outlined in previous research \cite{diwi,recbi}. To be specific, since Ochiai \cite{abreu2007accuracy} perform well, \ourSol employs the Ochiai Equation
$
score(s) = \frac{ef_s}{\sqrt{(ef_s + nf_s)(ef_s + ep_s)}}
$
to calculate the suspicious score of each statement. In the Equation, $ef_s$ and $nf_s$ represent the number of failing tests that execute and do not execute statement $s$, and $ep_s$ represents the number of passing tests that execute statement $s$.
In \ourSol, since there is only one given failing test program, the number of failing tests that execute statement $s$ ($ef_s$) is fixed at 1. Additionally, \ourSol focuses solely on the statements executed by the given failing test program, implying that the number of failing tests that do not execute statement $s$ ($nf_s$) is 0. As a result, the Ochiai Equation is simplified to:
$
score(s) = \frac{1}{\sqrt{(1 + ep_s)}}
$.

Once the suspicious score of each statement is obtained, \ourSol proceeds to calculate the suspicious score of each compiler file. Similar to previous studies \cite{diwi,recbi}, \ourSol aggregates the suspicious scores of the statements executed by the given failing test program within a compiler file to determine the suspicious score of the file
$
SCORE(f) = \frac{\sum_{i = 1}^{n_f} score(s_i)}{n_f}
$,
where the number of statements that the failing test program executes in the compiler file $f$ is denoted as $n_f$. \ourSol utilizes this information to calculate the suspicious score of each compiler file. By arranging the compiler files based on their suspicious scores in descending order, \ourSol yields a ranking list.

\subsubsection{Large Language Models (LLMs)} 
Recently, pre-trained Large Language Models (LLMs) such as ChatGPT \cite{chatgpt} become ubiquitous and have exhibited remarkable performance in numerous tasks, such as machine translation \cite{machine-translation}, text summarization \cite{text-sum}, classification \cite{classification} and code generation \cite{evalplus}.
Technically, LLMs can be directly employed to tackle specific downstream tasks by providing the task description to the model, which is known as prompt, without fine-tuning on specialized datasets. This is achieved through a technique known as prompt engineering \cite{DBLP:journals/csur/LiuYFJHN23, white2023prompt,reynolds2021prompt,yang2024prompt}. Prompt engineering aims to find the prompt that yields the best performance on specific tasks. Prior studies show that prompt engineering can achieve state-of-the-art performance on various downstream tasks \cite{brown2020language,niu2023empirical}. Benefiting from the huge potential of LLMs, there are increasing recent works showing that LLMs can be used for solving different tasks in software engineering \cite{xia2023automated,reynolds2021prompt,repairllm,deng2023large,liu2022fill}. In this study, we aim to unleash the power of LLMs in the field of test program generation tasks.

Many existing LLMs adopt the decoder of the Transformer architecture~\cite{vaswani2017attention}. Given a prompt containing the task description, the decoder generates the test programs $Y$ as a sequence of tokens, token-by-token, by following Equation~\ref{eq:gen},
\begin{equation}
    y_t = \argmax_{y} P(y | p, y_{<t})
    \label{eq:gen}
\end{equation}
where $y_t$ is the current token to be predicted, $y_{<t}$ refers to all the previously predicted tokens, and $p$ is the input prompt. The equation states that the current token $y_t$ to be predicted is determined by selecting the token $y$ that maximizes the conditional probability $P(y | p, y_{<t})$, given the prompt $p$ and the previously generated tokens $y_{<t}$.
Because a generated test program $Y$ depends on the input prompt $p$ and the search space for the input prompt $p$ is huge, finding the best prompt $p$ to generate the effective test program $Y$ can be challenging. In this work, we propose \ourSol to automatically find the prompt $p$ that can generate more effective test programs $Y$ for the compiler bug isolation task.

Two main categories of LLMs are available to the community for code generation tasks: infilling and general \cite{evalplus,deng2023large,cassano2023multipl}. Infilling models (e.g., CodeGen \cite{codegen}, Incoder \cite{incoder}, and PolyCoder \cite{polycoder}) are used to insert the most natural code based on bi-directional context (e.g., in the middle of a code snippet), while general models (e.g., LLaMA \cite{llama}, Alpaca \cite{alpaca}, ChatGPT \cite{chatgpt}, Vicuna \cite{vicuna}, and GPT4ALL \cite{gpt4all}) target to generate a complete code snippet given the left context only by a natural language description. In this study, we consider general models mainly due to the fact that general models follow the prompt-response dialog paradigm, which involves minimal human effort and fits our objective.

\begin{figure}[t]
\centering
\begin{minipage}{.22\textwidth}
\centering 
\begin{lstlisting}[caption={faling test program},basicstyle=\scriptsize]
short s;int a, b, c;
volatile int v;
static int u[] = {0,0,0,0,0,1};

void foo() {
  int i,j;
  for (; b <= 0; ++b) {
    int k; int d = 0;
    for (; d <= 5; d++) {
      int *l = &c;
      int e = 0;
      for (; e <= 0; e++) {
        int *m = &k;
        unsigned int n = u[d];
        i = !a ? n : n / a;
        j = s ? 0 : (1 >> v);
        *m = j;
      }
      *l = k < i;
    }
  }
}
int main() {
  foo();
  return 0;
}
$clang -O2 test.c ; ./a.out
$clang -O3 test.c ; ./a.out
$Floating point exception ...
\end{lstlisting}
\end{minipage}%
\hspace{1.5em}
\begin{minipage}{.22\textwidth}
\centering
\begin{lstlisting}[caption={passing test program},basicstyle=\scriptsize,escapechar=@]
short s;int a, b, c;
volatile int v;
static int u[] = {0,0,0,0,0,1};

void foo() {
  int i,j;
  for (; b <= 0; ++b) {
    int k; int d = 0;
    for (; d <= 5; d++) {
      int *l = &c;
      int e = 0;
      for (; e <= 0; e++) {
        int *m = &k;
        unsigned int n = u[d];
        i = !a ? n : n / a;
        j = s ? 0 : (1 >> v);
        *m = j;
        @\colorbox{light-gray}{\makebox(40,2){if (a==0) v = s;}}@
        @\colorbox{light-gray}{\makebox(55,2){else if(a>10) s=a+1;}}@
        @\colorbox{light-gray}{\makebox(26,2){else {s = 1;}}}@
      }
      *l = k < i;
    }
  }
}
int main() {
  foo();
  return 0;
}
\end{lstlisting}
\end{minipage}
\caption{LLVM bug \href{https://bugs.llvm.org/show_bug.cgi?id=16041}{\#16041} (the highlighted \colorbox{light-gray}{\makebox(35,4){gray code}} in (b) is generated by \ourSol)}
\label{fig:motivating-example}
\end{figure}

\subsection{Motivating Example} \label{sec:motivating-example}

\begin{figure*}[t]
\centering
\includegraphics[width=0.86\linewidth]{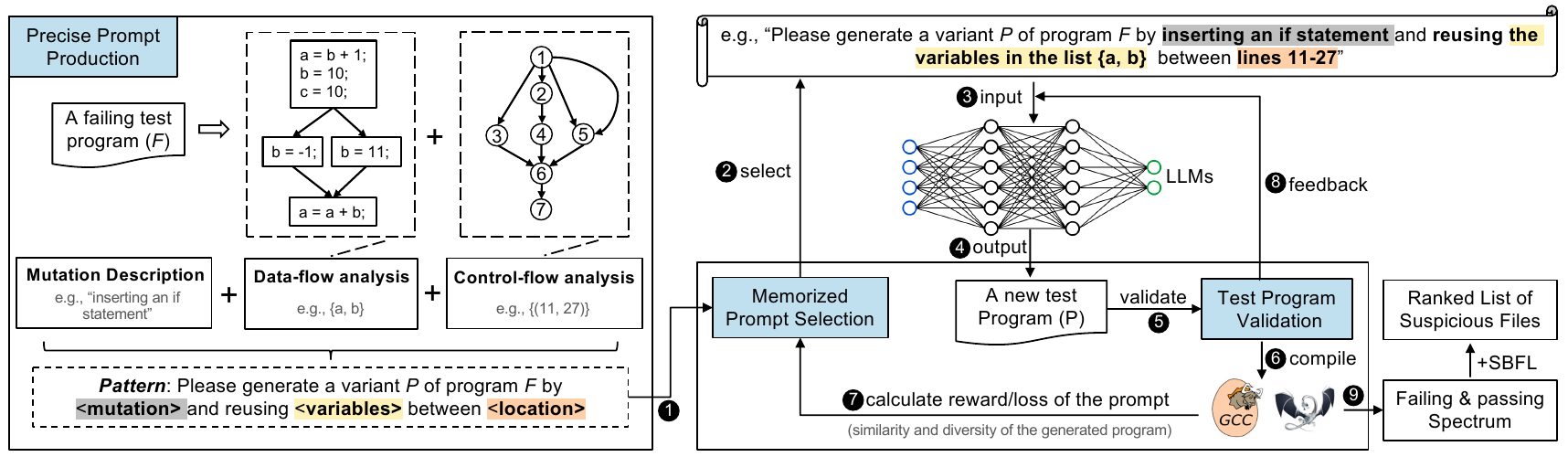}
\vspace{-1em}
\caption{Overview design of \ourSol}
\label{fig:overview}
\end{figure*}

Fig. \ref{fig:motivating-example}(1) showcases a failing test program that exposes a bug in the LLVM-3.4 compiler at the -O3 optimization level. The program introduces a {\tt division by zero} in the {\it induction variable elimination} optimization pass in LLVM, causing the miscompilation bug. Notably, Fig. \ref{fig:motivating-example}(2) represents a passing test program generated by our proposed solution (\ourSol), which can not trigger the bug in the LLVM compiler. Next, we show the limitations of existing approaches and the advantages of our approach in generating the passing test program.

\subsubsection{Limitations in Existing Approaches}
Existing approaches, such as DiWi \cite{diwi} and RecBi \cite{recbi}, face limitations in generating the above passing test programs due to three primary reasons.
First, the mutation strategies employed in DiWi and RecBi exhibit certain limitations. For instance, DiWi only supports local mutation operators that do not alter the control flow of the failing test program. Consequently, it cannot generate crucial elements such as the highlighted gray \texttt{if} statement ``if (a == 0)'' shown in Fig. \ref{fig:motivating-example}(2). Similarly, RecBi allows the insertion of structural conditions, like ``if (a == 0)'', but lacks the capability to generate the corresponding statement bodies, such as ``{s = v;}''. Moreover, both DiWi and RecBi rely on a random selection strategy for determining which variables to use and where to insert them during the mutation process, leaving the transformation outcomes largely dependent on chance.
Second, the mutation process involved in existing approaches necessitates substantial human effort. Before mutation, additional code must be written to collect essential contextual information (e.g., the names of variables like \texttt{s, a, b, c} and their defined types) and extract relevant elements (e.g., \texttt{if} statements) from the existing test programs. Manual coding is required to randomly determine suitable locations for inserting the new code snippets into the failing test programs during the mutation. Third, they pay little attention to the validity of the generated test program, which can have the side effect of compiler bug isolation.
In summary, the aforementioned limitations underscore the ineffectiveness and high demand of human efforts of existing approaches in generating high-quality test programs. These challenges highlight the need for a new solution that overcomes these shortcomings and effectively generates passing test programs.

\subsubsection{Advantages of Our Approach} 
Compared to DiWi \cite{diwi} and RecBi \cite{recbi}, \ourSol excels in generating effective passing test programs by taming the capabilities in LLMs. First, a precise prompt such as ``{\it Please generate a variant program P of the input program F by inserting an if statement and reusing the variables in the list \{a, s, v\} between lines 12-18}'' is produced to guide LLM in mutating the given program following certain requirements. Instead of using a vague prompt, \ourSol considers more detailed information in the program that can increase the likelihood of flipping from failing to passing to construct a precise prompt. In this way, the new test program is generated by \ourSol via inserting the new {\tt if}-statement shown in gray in Fig. \ref{fig:motivating-example}(2). Notably, \ourSol supports the insertion of bodies (such as ``{s = v;}'') in structural mutation, which increases the diversity of the generated test program. Second, benefiting from the LLMs' prompt-response dialog paradigm, the whole mutation process only involves little human effort compared with previous studies.
In addition, \ourSol detects and filters potential invalid test programs that contain undefined behaviors, further boosting the capabilities of \ourSol in terms of compiler bug isolation.

\section{The design of \ourSol} \label{sec:approach}

{\bf Overview.} 
Fig. \ref{fig:overview} illustrates the general workflow of \ourSol, which addresses the two main challenges and leverages the capabilities of LLMs to generate effective test programs for compiler bug isolation. \ourSol first generates precise prompts and collects all the generated prompts (\circled{1}) in the {\bf precise prompt production component}. Then, \ourSol selects a prompt (\circled{2}) via a {\bf memorized prompt selection component} and utilizes it as input for the LLMs (\circled{3}). Next, the \ourSol produces a new test program (\circled{4}), which undergoes a {\bf test program validation component} (\circled{5}). If the generated program is valid, it is compiled (\circled{6}), and coverage information of compiler source files is collected. Also, the quality of the generated test program will be measured with similarity and diversity metrics in (\circled{7}), which serve as the input of the memorized prompt selection component to help select better prompts. However, if the program is invalid due to semantic errors, \ourSol provides the feedback prompts (\circled{8}) to the LLMs, guiding them not to make the same mistakes again. Ultimately, upon reaching the termination condition (e.g., 1 hour), \ourSol employs SBFL along with the failing and passing spectra to rank suspicious files (\circled{9}). Specifically, the precise prompt production component is designed to address the first challenge of the 
formulation of precise prompts, and two other components, i.e., the memorized prompt selection component and lightweight test program validation component, are utilized to tackle the second challenge of selecting specialized prompts. We provide further details regarding this in the subsequent sections.

\subsection{Precise Prompt Production}
This section presents the precise prompt production component, which addresses the first challenge of the formulation of precise prompts. 
We first outline the design of the prompt production pattern for LLMs and then show how to utilize program complexity metrics, e.g., data-flow and control-flow complexity, to populate the pattern.

\subsubsection{Prompt Pattern for Program Mutation}

The following is the pattern designed in \ourSol for constructing effective prompts for LLMs. 

\begin{framed}
\vspace{-0.5em}
\noindent
\textbf{\emph{Pattern:}} Please generate a variant program {\it \bf P} of the input program {\it \bf F} by \colorbox{light-gray}{\makebox(70, 5){\textless mutation rule \textgreater}} and reusing the \colorbox{light-green}{\makebox(52, 5){\textless variables \textgreater}} between \colorbox{light-blue}{\makebox(46, 5){\textless location \textgreater}}.
\vspace{-0.5em}
\end{framed}
In the pattern, {\it \bf P} refers to the newly generated test program, and {\it \bf F} refers to the given failing test program. In the rest,
\colorbox{light-gray}{\makebox(70, 5){\textless mutation rule \textgreater}} means the actual mutation operation; \colorbox{light-green}{\makebox(52, 5){\textless variables \textgreater}} and \colorbox{light-blue}{\makebox(46, 5){\textless location \textgreater}} describe the specific requirements when conducting the mutation. 

We reuse the existing mutation rules (cf Table \ref{tab:mutation-rule}) in the pattern. Since compilers use different strategies to optimize the programs that have different data or control flow \cite{csmith},
our intuition is that,  instead of randomly mutating programs in existing approaches, \textit{mutating the most complex part of the failing program is more likely to flip the failing into passing}. To this end, \ourSol measures the most complex part by two means: the variable that holds complex data flow and the location that involves complex control flow. 

\subsubsection{Data-flow Complexity-guided Variable Selection}

The data-flow analysis aims to output the most complex variables defined and used in the given failing test program.

Due to the fact that we aim to investigate the complexity of a variable by examining how this variable can affect and to what extent,
we deem that the more a variable is defined (or assigned), the more complex dependence is held on the variable, thus contributing to the data-flow complexity. 

Following the existing definition of data-flow complexity \cite{mccabe1976complexity}, 
we follow the existing work \cite{newman2016srcslice} and opt for the variable {\it def-use} chain \cite{compiler-book} to analyze the data-flow complexity of a program.
Specifically, we calculate the complexity of a variable ($Comp_{var}$) by the following Equation:
\begin{equation}
Comp_{var} = N_{def} + N_{use}
\end{equation}
where the $N_{def}$ counts the number of times that a variable is defined, including redefining or assigning. Note that {\it Def} keeps track of the changes in a variable, so the data dependency analysis is included.
$N_{use}$ counts the number of times that a variable is used. This refers to the value of a variable value being used in some computation with no modification to the variable’s value.
Under the above Equation, the data flow analysis upon the failing test program in Fig. \ref{fig:motivating-example}(1) will produce a variable list {\tt \{a, s, v\}}, which represents the Top-3 complex variables used in the program.

\subsubsection{Control-flow Complexity-guided Location Selection}

The aim of the control-flow analysis is to output the location of the most complex statement in the program.
We leverage the control-flow graph of the input failing program, where each node represents the statement and the edge indicates the execution flow.
When the Control-Flow Graph (i.e., CFG) is available, we calculate the complexity of each statement using the following Equation (3) based on cyclomatic complexity \cite{mccabe1976complexity}:
\begin{equation}
    Comp_{control} = N_{edge} - N_{node} + 2
\end{equation}
where $N_{edge}$ and $N_{node}$ represent the number of edges and nodes in the CFG, respectively. Since the cyclomatic complexity is not designed to measure the complexity at the statement level, we count the complexity of each statement by obtaining the complexity values during cyclomatic complexity calculation. Note that we intentionally ignore the statement that includes the oracle (i.e., having the \texttt{printf} or \texttt{abort} function). The reason is that changing the code block, including the test oracle, is more likely to break the oracle, meaning a fake passing test program is probably generated \cite{recbi,diwi}. 
Taking the code example shown in Fig. \ref{fig:motivating-example}(1) again, the control-flow analysis designed in \ourSol will indicate that the most complex control flow lies in the for loop between Lines 12 - 18.
In this way, after getting the variable list and the desired location to be inserted, \ourSol generates certain prompts based on the designed pattern. For example, one of the generated prompts is:
\begin{framed}
    \vspace{-0.5em}
    \noindent
\emph{``Please generate a variant program P of the input program F by \colorbox{light-gray}{\makebox(110, 5){\textless inserting an if statement \textgreater}} and reusing the \colorbox{light-green}{\makebox(120, 5){\textless variables in the list \{a, s, v\} \textgreater}} between \colorbox{light-blue}{\makebox(55, 5){\textless lines 12-18 \textgreater}}''}.
\vspace{-0.5em}
\end{framed}

As aforementioned in Section \ref{sec:introduction},  not every prompt contributes equally to a specific failing test program. Furthermore, LLMs may make different mistakes when mutating a program: e.g., LLMs may incur a syntax error in a failing test program but a semantic error in another program. Giving all the error information as feedback prompts is not necessary.  Hence, a specialized prompt selection strategy is supported and developed in \ourSol.

\begin{table}[t!]
	\footnotesize
		\centering
		\caption{Mutation rules applied in the prompt pattern}
		\vspace{-1em}
		\begin{tabular}{clcr} 
			\toprule
			\textbf{ID}& \textbf{Description of Rules} \\
			\midrule
                1   & inserting an if statement  \\
                2 & inserting a loop (i.e., while or for) statement   \\
                3   & inserting a function call  \\
                4   & inserting a goto statement \\
			5   & inserting a qualifier (i.e., volatile, const, and restrict) \\
                6   & removing a qualifier (i.e., volatile, const, and restrict) \\
			7 & inserting a modifier (i.e., long, short, signed, and unsigned)   \\
                8   & removing a modifier (i.e., long, short, signed, and unsigned)  \\
                9   & replacing a constant with another valid one  \\
			10 & replacing a binary operator with another valid one   \\
                11  & removing a unary operator on the variables  \\
                12   & replacing a unary operator on the variables  \\
			13 & replacing a variable  with another valid one  \\
              
			\bottomrule
			\label{tab:mutation-rule}
		\end{tabular}
		\vspace{-1em}
\end{table}

\subsection{Memorized Prompt Selection}
This subsection and the next subsection present memorized prompt selection and lightweight test program validation components to address the second challenge of selecting the specialized prompts.
In this subsection, we first give the background of reinforcement learning and then detail the memorized prompt selection. 

\subsubsection{Reinforcement Learning}

Reinforcement learning (a kind of memorized search) is a field of study that aims to teach an agent how to take actions within an environment to maximize its cumulative reward over the long term \cite{kaelbling1996reinforcement,sutton2018reinforcement}. 
Key roles in reinforcement learning include (1) {\bf Agent}: the role of a learner and decision-maker; (2) {\bf Environment}: everything that is composed of and interacts with something other than the agent; (3) {\bf Action}: the behavioral representation of the agent body; (4) {\bf State}: the information that the capable body obtains from the environment; (5) {\bf Reward}: feedback from the environment about the action.
Reinforcement learning can generally be classified into value-based algorithms (e.g., Deep Q Learning algorithm \cite{mnih2013playing}) and policy-based algorithms (e.g., Policy Gradients algorithm \cite{sutton1999policy}).

With the advancement of reinforcement learning, a class of algorithms known as Actor-Critic (AC) algorithms have been proposed \cite{konda1999actor}, combining elements of value-based and policy-based strategies. 
In this study, we employ the Advantage Actor-Critic (i.e., A2C \cite{mnih2016asynchronous}) framework to address the challenge of compiler bug isolation, as it is both effective (in comparison to AC \cite{konda1999actor}) and suitable for single-thread and multi-thread systems (compared to Asynchronous Advantage Actor-Critic, i.e., A3C \cite{mnih2016asynchronous}).
In this study, \ourSol adopts the A2C (Advantage Actor-Critic) framework \cite{mnih2016asynchronous} to learn the effects of a prompt, enabling the generation of effective passing test programs for a specific compiler bug. We opt for the A2C framework mainly because it has demonstrated practical effectiveness, efficiency, and stability with low variance \cite{grondman2012survey}, making it a suitable choice.

\subsubsection{Reinforcement Learning for Prompt Selection}

The effectiveness of randomly applying mutations upon a given failing test program can be limited \cite{recbi}. 
Consequently, we do not randomly select prompts for LLMs to generate program variants; we need to efficiently generate more effective passing test programs within a given time tailored to a specific compiler bug.
To accomplish this goal, we employ reinforcement learning in \ourSol, where the quality of the generated passing test programs serves as the reward metric of a prompt. 
In this study, every prompt is composed of the input code, the instance of the prompt pattern (including the mutation operator, and the most complex variables and lines). As shown in Fig. 3, the precise mutation info is fed to and selected by the reinforcement learning model. For example, ``inserting an if statement and reusing the variables in the list {a, s, v} between lines 12-18'' and ``inserting a loop statement and reusing the variables in the list {a, s, v} between lines 12-18'' can be two different mutation info of the prompt (there are 13 in total based on the mutation operators listed in Table 1). In short, our reinforcement learning model learns how to select the optimal mutation operator, i.e., the one that can be the most effective one for producing high-quality witness test programs.

\begin{figure}[t]
\centering
\includegraphics[width=0.85\linewidth]{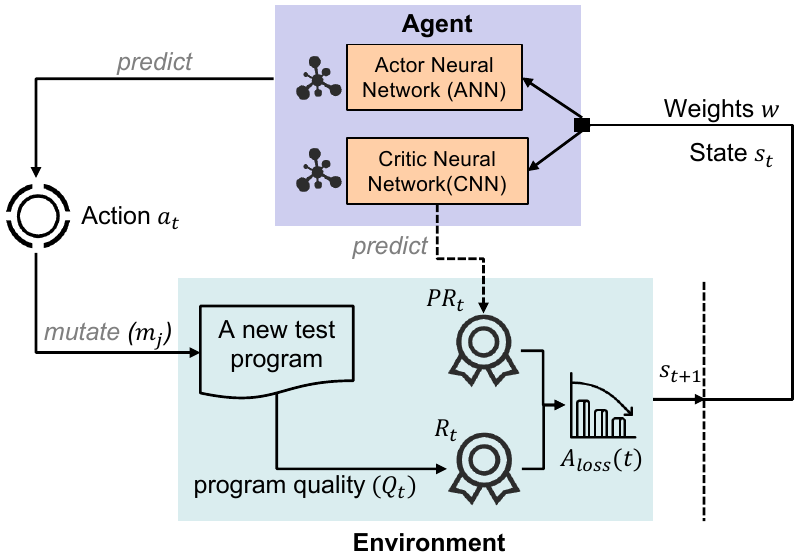}
\begin{tablenotes}
        \footnotesize
        \item * $PR_t$: potential reward at time $t$
        \item * $R_t$: actual reward at time $t$
        \item * $A_{loss}$: advantage loss function at time $t$
        \item * $m_{j}$: a selected prompt ($j \in \{1, 13\}$)
\end{tablenotes}
\caption{Memorized prompt selection guided by reinforcement learning: The agent uses an ANN to predict an action $a_t$ (i.e., a prompt $m_j$ used for mutating a program). After obtaining the new test program from LLMs, the environment calculates the actual reward $R_t$ based on the quality ($Q_t$) of the newly generated test program. In the meantime, CNN in the agent predicts a potential reward $PR_t$. Next, the advantage loss ($A_{loss}$) is measured by combining the actual reward and the potential reward. Later, the weights ($w$) of ANN and CNN and state ($s_t$) will be updated to the agent and help the agent generate a better prompt.}
\label{fig:drl}
\end{figure}

Fig.~\ref{fig:drl} provides an overview of the reinforcement learning-based strategy for selecting prompts in \ourSol.
Following the A2C framework, \ourSol first initializes two neural networks: the Actor Neural Network (ANN) and the Critic Neural Network (CNN) in the agent. The ANN predicts the probability distribution of actions based on historical knowledge, enabling the subsequent selection of an optimal action by \ourSol.
CNN predicts the potential reward that can be accumulated from the current state to a future state after executing the selected action, incorporating future knowledge.
\ourSol chooses an action $a_t$ to select a prompt (randomly chosen for the first time) and measure the quality ($Q_t$) of the newly generated test program.
To facilitate learning, \ourSol employs an advantage loss function ($A_{loss}$) based on the predicted potential reward ($PR$) and the actual reward ($R$) obtained from the selected prompt. Finally, \ourSol updates the status of all the ANN, CNN, and states to the agent.
\ourSol repeatedly selects a prompt to generate test programs until the termination condition (e.g., 1-hour limit is reached or 10 program variants are generated) is reached.

Consistent with existing A2C-based approaches \cite{konda1999actor,mnih2016asynchronous}, both ANN and CNN in \ourSol comprise a single hidden layer to ensure lightweight and fast convergence. 
The following provides detailed explanations of the most important parts, including the actual reward and advantage loss function calculations.

\smallskip
\noindent
\textbf{Measuring Actual Reward.}  An essential factor contributing to the effectiveness of the A2C-based approach lies in determining the actual reward subsequent to applying a prompt.
Drawing the inspiration from previous studies \cite{diwi,recbi}, a collection of proficient passing test programs needs to meet both {\it similarity} and {\it diversity} criteria. The {\it similarity} entails that each passing test program should exhibit a comparable compiler execution trace to that of the given failing test program. Consequently, following the principles of SBFL, the suspicion associated with a greater number of buggy-free files can be diminished. The {\it diversity} requires that different passing test programs possess distinct compiler execution traces from one another to reduce suspicion regarding various buggy-free files. By doing so, aggregating a set of passing test programs that have undergone mutation facilitates the effective isolation of genuinely faulty files by circumventing bias. Both {\it similarity} and {\it diversity} rely on the {\it distance} metric, which is defined as:
\begin{equation}
    dist(a,b) = 1 - \frac{Cov_a \cap Cov_b}{Cov_a \cup Cov_b} \label{formu:dist}
\end{equation}
where the $dist(a,b)$ represents the coverage distance between two test programs $a$ and $b$, which is determined using the Jaccard Distance measurement \cite{diwi}. $Cov_a$ and $Cov_b$ represent the sets of statements covered in compilers by test programs $a$ and $b$, respectively.

Denoting the set of generated passing test programs as $p = \{p_1, p_2, ..., p_n\}$ and the failing test program as $f$, we formalize the metrics of {\it similarity} and {\it diversity} achieved by the set of passing test programs, represented as Equation \ref{formu:sim} and Equation \ref{formu:div}, respectively.

\begin{equation}
    sim = \frac{\sum_{i = 1}^{N} (1 - dist(p_i, f))}{N}  \label{formu:sim}
\end{equation}
\begin{equation}
    div = \frac{\sum_{i = 1}^{N - 1}\sum_{j = i+1}^{N}dist(p_i, p_j)}{N(N-1)/2}  \label{formu:div}
\end{equation}
where $N$ is the number of passing test programs.

At time step $t$, once a passing test program is generated, \ourSol evaluates the effectiveness of the current set of passing test programs by linearly combining the attained measures of {\it similarity} and {\it diversity} in Equation \ref{formu:quality}. 
\begin{equation}
    Q_t = n (\alpha \times div_t + (1 - \alpha)\times sim_t)  \label{formu:quality}
\end{equation}
where the coefficient $\alpha$ represents the weighting factor for the linear combination of {\it diversity} and {\it similarity} in Equation \ref{formu:quality}. Additionally, following the existing study \cite{recbi}, Equation \ref{formu:quality} also incorporates another coefficient $n$, which corresponds to the size of the set of passing test programs.

Subsequently, \ourSol determines whether to accept the generated passing test program based on whether it can enhance the overall quality of the set of passing programs compared to the previous time step denoted as $t-1$. The Equation \ref{formu:delta} outlines the computation of the enhanced quality relative to the previous time step.
\begin{equation}
    \triangle Q_t = Q_t - Q_{t-1} \label{formu:delta}
\end{equation}

Nevertheless, in each state, only one prompt is chosen to generate a passing test program, and the performance of a prompt can vary significantly depending on different bugs.
To balance the influence of the diverse performance of prompts, \ourSol incorporates both the current time step improvement and the historically accumulated improvement attributed to the current prompt. This combined value serves as the actual reward obtained at the current time step, which is defined as follows:
\begin{equation}
    R_t = \frac{\sum_{i = 1}^{t} \triangle Q_i}{T(m_j)}  \label{formu:reward}
\end{equation}
Here, $T(m_j)$ represents the count of times the $m_j$ (a prompt) has been chosen to mutate the failing test program.
$\triangle Q_i$ is defined as zero ($\triangle Q_i$ = 0) if the selected prompt at the $i^{th}$ time step is not $m_j$; 
if the selected prompt is $m_j$, $\triangle Q_i$ is computed using Equation \ref{formu:delta}. Whether selecting a new prompt $m_j$ or previously selected prompt $m_j$ depends on the decision made by ANN in the agent.

\smallskip
\noindent
\textbf{Calculating Advantage Loss.}
While the actual reward ($R_t$) is obtained at the current time step $t$, \ourSol also utilizes CNN to predict the potential reward ($PR_{t+i}$).
To effectively consider future factors, A2C incorporates an advantage loss function. This function is designed to address the issue of high variance in the two neural networks and prevent convergence towards local optima \cite{mnih2016asynchronous}. The advantage loss function is expressed as follows in Equation \ref{formu:loss}:
\begin{equation}
    A_{loss}(t) = \sum\limits_{i=t}^{t+u}(\gamma^{i-t} R_i) + \gamma^{u+1}PR_{t+u}-PR_t  \label{formu:loss}
\end{equation}
the variable $u$ indicates that the CNN considers the future $u$ consecutive states and actions when predicting the potential reward. $\gamma$ represents the weight assigned to the actual future reward. $PR_{t+u}$ and $PR_t$ denote the predicted potential rewards at the $(t+u)^{th}$ and $t^{th}$ time steps, respectively, as determined by CNN. Notably, \ourSol repeats this process for $u$ times within a time step to approximate the actual future reward.

Using the loss computed by the advantage function in Equation \ref{formu:loss}, \ourSol proceeds to update the weights of both the ANN and CNN following Equation \ref{formu:weight}.
\begin{equation}
    w = w + \beta\frac{\partial (\log P_w(a_t|s_t)A_{loss}(t))}{\partial w}  \label{formu:weight}
\end{equation}
where $s_t$ represents the current state, while $a_t$ denotes the corresponding action. $P_w(a_t|s_t)A_{loss}(t)$ represents the probability of performing action $a_t$ at state $s_t$ based on the parameters $w$ in both the ANN and CNN. $\beta$ represents the learning rate of the weight updates.

\subsection{Lightweight Test Program Validation}

Existing studies \cite{diwi,recbi} pay little attention to the validity of the mutated test programs. First, the test programs generated by these approaches may contain undefined behaviors. As demonstrated in our evaluation results in Section \ref{sec:answer2}, such test programs reduce the effectiveness of compiler bug isolation. Second, existing approaches may generate test programs that do not have a test oracle, which can also affect the effectiveness of bug isolation.
Third, existing studies are unaware of the errors they made during the mutation process, so they frequently make the same mistakes when mutating test programs.

To address the above limitations,
\ourSol designs a semantic validation to filter away semantically invalid test programs and utilizes a test oracle validation to fix the test programs that violate test oracles. 
Additionally, \ourSol collects all the validation errors in the test programs generated by LLMs and gives such information as feedback prompts to LLMs to avoid repeating the same mistakes.
Next, we detail the validation processes.

\subsubsection{Program Semantic Validation}
We choose static analysis checks on the newly generated test program that may contain any kinds of undefined behaviors, which proved to be lightweight compared with dynamic analysis approaches \cite{framac}. As shown in step \circled{5} in Fig. \ref{fig:overview}, the newly generated test program is transferred to this component for checking the semantic validity. 
Based on the analysis capabilities from Frama-C\cite{framac}, we opt for five different categories of undefined behaviors in \ourSol:
\begin{itemize}[leftmargin=1em,nosep]
    \item \texttt{mem\_access}: invalid memory access, such as out-of-bound read or out-of-bound write.
    \item \texttt{shift}: invalid RHS (Right-Hand Shift) or LHS (Left-Hand Shift) operand for right or left shift operation.
    \item \texttt{index\_bound}: accessing out-of-bounds index of an array.
    \item \texttt{initialization}: accessing uninitialized left-value, i.e., use a variable before it was uninitialized.
    \item \texttt{division\_by\_zero}: a number is divided by zero.
\end{itemize}

If one of the above kinds of undefined behaviors is detected, the semantic error information will be updated to LLMs to avoid repeating the same mistakes. For example, if a new test program contains an undefined \texttt{divided by zero} behavior, an additional prompt, ``\textit{The above program contains a kind of undefined behavior \texttt{divided by zero}, please do not generate such test programs again.}'', will be fed to LLMs in step \circled{8}.

\subsubsection{Test Oracle Validation}
\smallskip
\noindent
\textbf{Test Oracles}.
It is also required to check whether a generated test program is passing or failing \cite{diwi,recbi,zhou2022locseq,yang2022isolating}. According to the types of compiler bugs (i.e., crash bugs and wrong-code bugs), following the existing works \cite{diwi,recbi}, \ourSol considers two types of test oracles:
(1) Regarding crash bugs (i.e., the compiler crashes when using some compilation options to compile a test program), the test oracle is whether the compiler still crashes when using the same compilation options to compile a generated test program.
(2) Regarding wrong-code bugs (i.e., the compiler mis-compiles a test program without any failure messages, causing the test program to have inconsistent execution results under different compilation options), the test oracle is whether a generated test program still produces inconsistent execution results under the compilation options producing the previous inconsistencies.
Similar to DiWi \cite{diwi} and RecBi \cite{recbi}, LLMs may not put in the test oracles in a generated test program; so, we apply another
heuristic
validation to check for missing oracles.
Specifically, we check if the generated test program contains the same number of \texttt{abort} or \texttt{printf} statements as the given failing test program.

If a newly generated test program passes the above validations,
the buggy compiler is used to compile the generated test program in step \circled{6}. 
\ourSol collects the semantic error or test oracle information and uses it as feedback prompts to guide LLMs not to generate programs with the same kinds of errors. 
It is worth noting that since different test programs may contain different kinds of errors, the feedback prompts could be different.

\section{Evaluation} \label{sec:evaluation}

\begin{table*}[t]
\renewcommand{\arraystretch}{1.0}
\centering
\scriptsize
\setlength{\tabcolsep}{4.5pt}
\caption{Compiler bug isolation effectiveness comparison with two state-of-the-art approaches (under Setting-1 in RQ1)}
\vspace{-1em}
\begin{threeparttable}
\begin{tabular}{c|c|cc|c c|c c|c c|c c|c c}
\toprule 
\multirow{2}{*}{\textbf{Subject}} &\multirow{2}{*}{\textbf{Approach}} & \textbf{Num.} &\textbf{$\Uparrow_{Top-1}$} & \textbf{Num.}& \textbf{$\Uparrow_{Top-5}$} & \textbf{Num.}& \textbf{$\Uparrow_{Top-10}$} & \textbf{Num.} & \textbf{$\Uparrow_{Top-20}$} &\multirow{2}{*}{\textbf{MFR}} & \textbf{$\Uparrow_{MFR}$} &  \multirow{2}{*}{\textbf{MAR}} & \textbf{$\Uparrow_{MAR}$} \\

& &  \textbf{Top-1} & (\%) & \textbf{Top-5 }&  (\%) & \textbf{Top-10 }&  (\%) & \textbf{Top-20} & (\%) &   & (\%) &  & (\%)\\
\midrule 
\multirow{3}{*}{ \textbf{GCC} }
 & DiWi \cite{diwi}   &5.60&	66.07	&19.90	&25.13&	31.30&	16.29&	41.70	&7.43	&22.57&	29.83&	23.10&	29.19 \\
 & RecBi \cite{recbi} &7.70&	20.78&	23.90&	4.18	&34.20&	6.43&	42.50&	5.41&	20.53&	22.84&	21.06&	22.33 \\
  & \ourSol           &9.30	&-	&24.90 &-	&36.40	&-	&44.80	&-	&15.84&	-	&16.35&	- \\
\midrule \multirow{3}{*}{ \textbf{LLVM} }
 & DiWi \cite{diwi}   &4.30&	74.42&	19.20	&18.23&	29.20	&17.81	&40.00&	20.00	&27.42	&43.97	&27.50	&43.82 \\
 & RecBi \cite{recbi} &5.80&	29.31&	19.80	&14.65	&30.00	&14.67	&42.40&	13.21	&25.43	&39.59	&25.56	&39.54 \\
  & \ourSol           & 7.50	&-	&22.70&	-	&34.40&	-	&48.00&	-	&15.36&	-	&15.45 & - \\
\midrule \multirow{3}{*}{ \textbf{ALL} }
  & DiWi \cite{diwi}   & 9.90	&69.70&	39.10&	21.74&	60.50	&17.02&	81.70&	13.59	&25.00	&37.58&25.30	&37.15 \\
  & RecBi \cite{recbi} & 13.50	&24.44	&43.70	&8.92&	64.20&	10.28	&84.90&	9.31&	22.98&	32.11&	23.31&	31.77 \\
  & \ourSol            & 16.80	&-	&47.60&	-	&70.80&	-	&92.80	&-	&15.60&	-	&15.90&	 - \\
 \bottomrule
\end{tabular}
 \begin{tablenotes}
        \footnotesize
        \item Note: Columns “$\Uparrow*$” present the improvement rates (\%) of \ourSol over the compared approaches in various metrics.
      \end{tablenotes}
  \end{threeparttable}
  \label{tab:rq1-1}
\end{table*}

\subsection{Experimental Setup}

\smallskip
\noindent
\textbf{Implementation of \ourSol.} \label{sec:implementation}
\ourSol was implemented utilizing OClint \cite{oclint} (v22.02), srcSlice \cite{newman2016srcslice} (v1.0), Gcov \cite{gcov} (v4.8.0), and PyTorch \cite{pytorch} (v1.10.1+cu113). OClint is served to calculate the cyclomatic complexity of each statement; srcSlice is used to get the data-flow complexity of the variables defined and used in the program; Gcov is applied for collecting compiler coverage information; PyTorch supports the A2C framework. For A2C, we set the hyperparameters with the default settings in the previous study \cite{recbi}. For the implementation of test program validation, we adopt Frama-C \cite{framac} (Phosphorus-20170501) to check the semantic validity of the test program, and we write Python scripts (python 3.8.5) to check if there are any test oracle violations. We adapt GPT-3.5 as the default LLMs in \ourSol, with the {\it temperature} parameter 1.0 in GPT-3.5 (see more detailed discussion on temperature settings in Section \ref{sec:discussion}).

\smallskip
\noindent
\textbf{Study Subjects}.
We use  GCC and LLVM as our subjects to assess the effectiveness of \ourSol. Both two compilers are widely used in existing literature \cite{tu2022detecting,jiang2021ctos,tu2022remgen,diwi,recbi,yang2022isolating,zhou2022locseq} and therefore constitute a comprehensive evaluation. On average, a GCC buggy version contains 1,758 files with 1,447K source lines of code (SLOC), while an LLVM buggy version comprises 3,265 files with 1,723K SLOC.

\smallskip
\noindent
\textbf{Benchmark}.
We utilize a benchmark consisting of 120 real compiler bugs, with an equal distribution of 60 GCC and 60 LLVM bugs, which includes all bugs studied in prior works \cite{diwi,recbi}. Specifically, each bug is accompanied by relevant buggy details, including the faulty compiler version, the failing test program, the buggy compilation options, and the faulty files which serve as the ground truth. 

\smallskip
\noindent
\textbf{Running Platform}. We conducted all the experiments on a workstation equipped with a 12-core CPU, Intel(R) Xeon(R) W-2133 CPU @ 3.60GHz, 64G RAM, and Ubuntu 18.04 operating system, without GPU support.

\smallskip
\noindent
\textbf{Evaluation Metrics.}
Each approach for isolating compiler bugs generates a list of suspicious compiler files. To evaluate the effectiveness of each approach, we measure the position of each buggy file in the ranking list with the help of the ground truth. In cases where multiple compiler files had the same suspicious scores, we follow the precedent set by prior studies \cite{jeffrey2008fault,pearson2017evaluating} and assign the worst ranking. Specifically, we compute the following metrics commonly used in compiler bug isolation studies \cite{diwi,recbi,zhou2022locseq,yang2022isolating}.

\begin{itemize}[leftmargin=1em,nosep]
    \item \textbf{Top-N}. This metric denotes the number of bugs that are effectively isolated and contained within the Top-N position, where N is a member of the set {1, 5, 10, 20} as specified in our study. A higher value of Top-N indicates a better performance of an approach.
    \item \textbf{Mean First Ranking (MFR)}. This metric represents the average rank of the first faulty file within the ranking list for each bug. The objective of MFR is to promptly isolate the initial defective element in order to expedite the debugging process. A smaller value is better, as it indicates that developers could localize the corresponding bug as quickly as possible.
    \item \textbf{Mean Average Ranking (MAR)}. This metric measures the average of the mean rank of every faulty file within the ranking list for each bug. The MAR metric is intended to isolate all faulty elements accurately. Similar to MFR, the approach with a smaller value of MAR is better. 
\end{itemize}

\subsection{Research Questions}

In this study, we aim to answer the following main research questions (RQs):
\begin{itemize} [leftmargin=1em,nosep]
    \item \textbf{RQ1}: Can \ourSol outperform state-of-the-art approaches (i.e., DiWi \cite{diwi} and RecBi \cite{recbi}) in terms of the effectiveness and efficiency of compiler bug isolation?
    \item \textbf{RQ2}: Can each main component, i.e., precise prompt production, memorized prompt selection, and lightweight test program validation, contribute to \ourSol?
    \item \textbf{RQ3}: Can \ourSol be easily extended with other LLMs for the compiler bug isolation task?
\end{itemize}

\subsection{Answers to RQ1} \label{sec:answer1}

\subsubsection{Experimental Settings}
We set up the following two settings to investigate RQ1:
\begin{itemize}[leftmargin=1em,nosep]
    \item Setting-1: terminate with the same running time (i.e., one hour). This is the standard comparison strategy used in existing compiler bug isolation studies \cite{recbi,diwi,zhou2022locseq,yang2022isolating} for evaluating their effectiveness.
    \item Setting-2: terminate when generating the same number (i.e., 10) of passing test programs. This setting aims to demonstrate the efficiency of \ourSol further. We opt for the number suggested by the existing empirical studies \cite{abreu2007accuracy}. We give the timeout of 2 hours if an approach cannot generate the desired number of test programs of a bug.
    We add this setting because during the experiments in Setting-1, we find the number of generated passing test programs by comparative approaches is different, and \ourSol could generate a larger number of passing test programs. Thus, whether the superior performance on \ourSol benefited from the more test programs or the quality of the newly generated programs is unknown. We conduct Setting-2 to investigate it further.
\end{itemize}

\smallskip
\noindent
\textbf{Comparison Strategies}.
For both settings, we repeatedly ran all the comparative approaches 10 times and calculated the average results of Top-N, MFR, and MAR metrics to reduce the influence of randomness. Specifically, we conduct the statistical test analysis followed by the suggestions from Arcuri and Briand \cite{arcuri2011practical,vargha2000critique}. Additionally, for Setting-2, we compare the execution time for each approach and speedups achieved by \ourSol.

\begin{table*}[t]
\renewcommand{\arraystretch}{1.0}
\scriptsize
\centering
\setlength{\tabcolsep}{4.5pt}
\caption{Compiler bug isolation effectiveness comparison with two state-of-the-art approaches (under Setting-2 in RQ1)}
\vspace{-1em}
\begin{threeparttable}
\begin{tabular}{c|c|cc|c c|c c|c c|c c|c c}
\toprule 
\multirow{2}{*}{\textbf{Subject}} &\multirow{2}{*}{\textbf{Approach}} & \textbf{Num.} &\textbf{$\Uparrow_{Top-1}$} & \textbf{Num.}& \textbf{$\Uparrow_{Top-5}$} & \textbf{Num.}& \textbf{$\Uparrow_{Top-10}$} & \textbf{Num.} & \textbf{$\Uparrow_{Top-20}$} &\multirow{2}{*}{\textbf{MFR}} & \textbf{$\Uparrow_{MFR}$} &  \multirow{2}{*}{\textbf{MAR}} & \textbf{$\Uparrow_{MAR}$} \\

& &  \textbf{Top-1} & (\%) & \textbf{Top-5 }&  (\%) & \textbf{Top-10 }&  (\%) & \textbf{Top-20} & (\%) &   & (\%) &  & (\%)\\
\midrule 
\multirow{3}{*}{ \textbf{GCC} }
 & DiWi \cite{diwi}   & 4.80	& 56.25& 	18.60& 	27.96& 	31.20	& 11.86	& 40.80	& 7.84	& 23.86& 	33.13& 	24.27	& 33.54 \\
 & RecBi \cite{recbi} & 6.80	& 10.29	& 23.10& 	3.03& 	34.20& 	2.05& 	42.80& 	2.80& 	18.81& 	15.19& 	19.28& 	16.34 \\
  & \ourSol           & 7.50	& -	& 23.80& 	-& 	34.90& 	-	& 44.00& 	-& 	15.96& 	-	& 16.13& 	- \\
\midrule \multirow{3}{*}{ \textbf{LLVM} }
 & DiWi \cite{diwi}   & 4.20&	88.10&	19.90	&11.56	&29.30	&17.06&39.40&	12.18&	26.13&	37.63&	26.22&	37.44 \\
 & RecBi \cite{recbi} & 6.10	&29.51	&20.60&	7.77&	32.90&	4.26&	41.30&	7.02&	19.56&	16.68&	19.64&	16.46\\
  & \ourSol           & 7.90	&-	&22.20&	-	&34.30&	-	&44.20	&-	&16.30&	-	&16.40	&- \\
\midrule \multirow{3}{*}{ \textbf{ALL} }
  & DiWi \cite{diwi}   & 9.00	&71.11&	38.50&	19.48&	60.50&	14.38&	80.20&	9.98&	25.00&	35.48&	25.24&	35.56 \\
  & RecBi \cite{recbi} & 12.90&	19.38&	43.70	&5.26	&67.10	&3.13&	84.10&	4.88&	19.19&	15.95&	19.46	&16.40 \\
  & \ourSol            & 15.40&	-	&46.00&	-	&69.20&	-	&88.20&	-	&16.13&	-	&16.27&	- \\
 \bottomrule
\end{tabular}
 \begin{tablenotes}
        \footnotesize
        \item Note: Columns “$\Uparrow*$” present the improvement rates (\%) of \ourSol over the compared approaches in various metrics.
      \end{tablenotes}
  \end{threeparttable}
  \label{tab:rq1-2}
\end{table*}

\subsubsection{Experimental Results}

\smallskip
\noindent
\textbf{Results for Setting-1}.
The comparison results of Setting-1 are presented in Table \ref{tab:rq1-1}. The first column represents subject compilers, and the second column shows different approaches. Columns 3-10 provide the Top-N metrics derived from the average values obtained from 10 runs of each approach, including the number of Top-N (i.e., {\bf Num. Top-N}) and improvement (i.e., \textbf{$\Uparrow_{Top-N}$ (\%)}) made by \ourSol. Columns 11-14 represent the {\bf MFR} (Mean First Ranking) and {\bf MAR} (Mean Average Ranking) metrics, as well as the improvement achieved by \ourSol.

Notably, \ourSol demonstrates its capability by successfully isolating 16.80, 47.60, 70.80, and 92.80 compiler bugs (out of a total of 120 compiler bugs in GCC and LLVM) within the Top-1, Top-5, Top-10, and Top-20 files, respectively. This accounts for the improvement of 69.70\%, 21.74\%, 17.02\%, and 13.59\% than DiWi and 24.44\%, 8.92\%, 10.28\%, and 9.31\% than RecBi, respectively.
Further analysis of the effects across different subject compilers revealed interesting findings. Despite the larger number of compiler files in LLVM compared to GCC, \ourSol exhibited slightly better results in the case of GCC. For instance, \ourSol achieved MFR and MAR values of 15.36 and 14.45 for GCC, while corresponding values for LLVM were 15.60 and 15.90. 

Compared with DiWi and RecBi, the evaluation reveals that \ourSol outperforms DiWi and RecBi across all metrics and for both GCC and LLVM compilers. Notably, \ourSol demonstrates significant improvement of 69.70\% and 21.74\% over DiWi in terms of Top-1 and Top-5. For RecBi, \ourSol could isolate 24.44\% and 8.92\% more bugs than RecBi in terms of Top-1 and Top-5.
In particular, the practical significance of the Top-5 metric is highlighted by previous research, which indicates that developers often discontinue the use of automated debugging tools if the faulty elements cannot be localized within the Top-5 positions. Consequently, \ourSol shows better practicality compared to DiWi and RecBi by substantially improving the effectiveness of compiler bug isolation, specifically in relation to the Top-5 metric.

Furthermore, in terms of MFR and MAR, \ourSol achieves a considerable improvement of 37.58\% and 37.15\% over DiWi and 32.11\% and 31.77\% over RecBi, respectively. The improvement of MFR and MAR demonstrates that \ourSol can promptly and accurately isolate more compiler bugs than DiWi and RecBi.

\smallskip
\noindent
\textbf{Results for Setting-2}.
From Table \ref{tab:rq1-2}, for all the 120 bugs, we can know that \ourSol also outperforms comparative approaches: \ourSol can isolate more bugs and hold the lowest MFR and MAR.

\begin{figure}[t]
\vspace{-1em}
\centering
\subfigure[Distribution of execution time]{
\begin{minipage}[t]{0.47\linewidth}
\centering
\includegraphics[width=4.3cm]{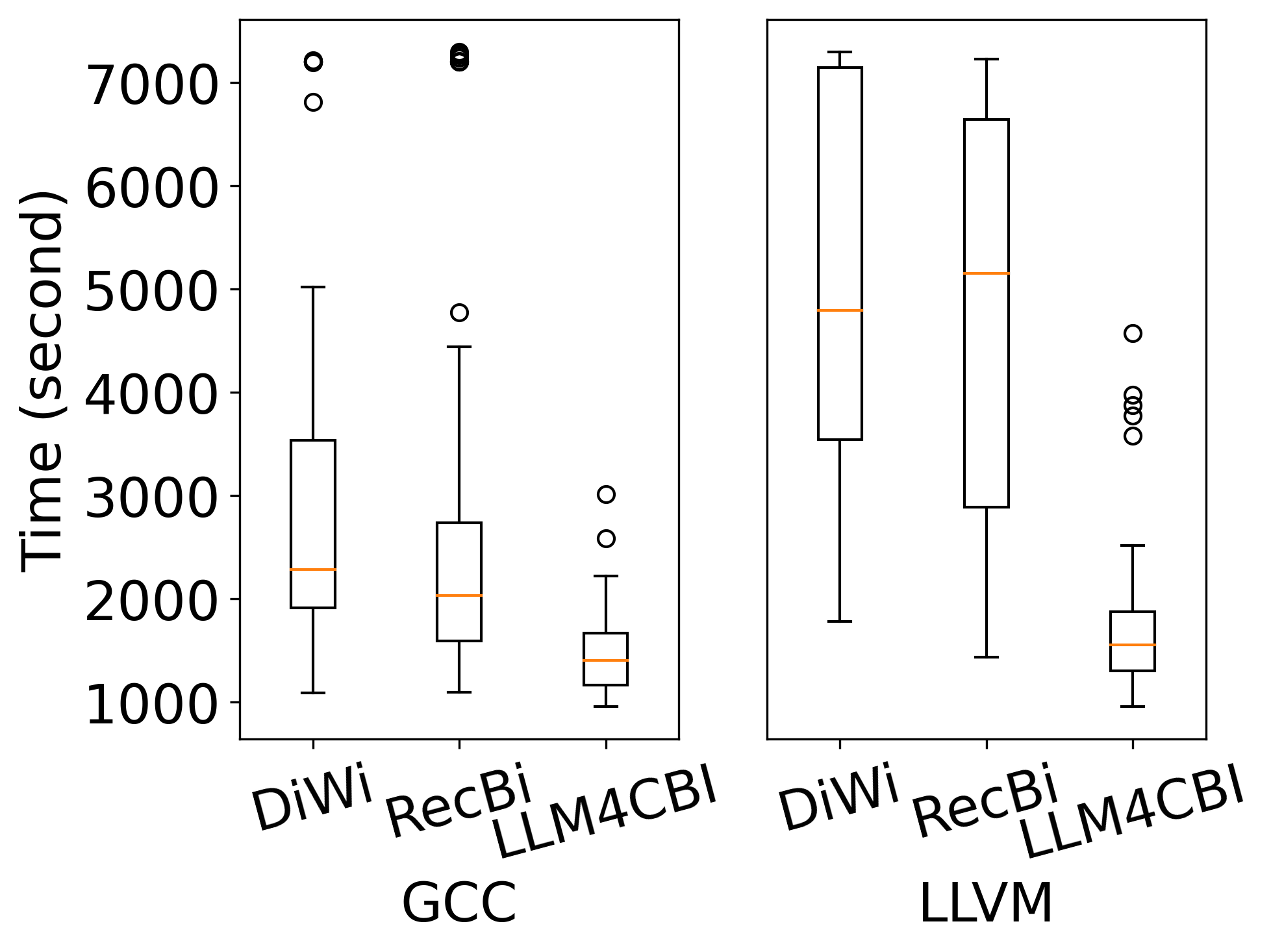}
\vspace{-1em}
\label{fig:time}
\end{minipage}
}
\subfigure[Speedups over DiWi and RecBi]{	
\begin{minipage}[t]{0.47\linewidth}
\centering
\includegraphics[width=4.4cm]{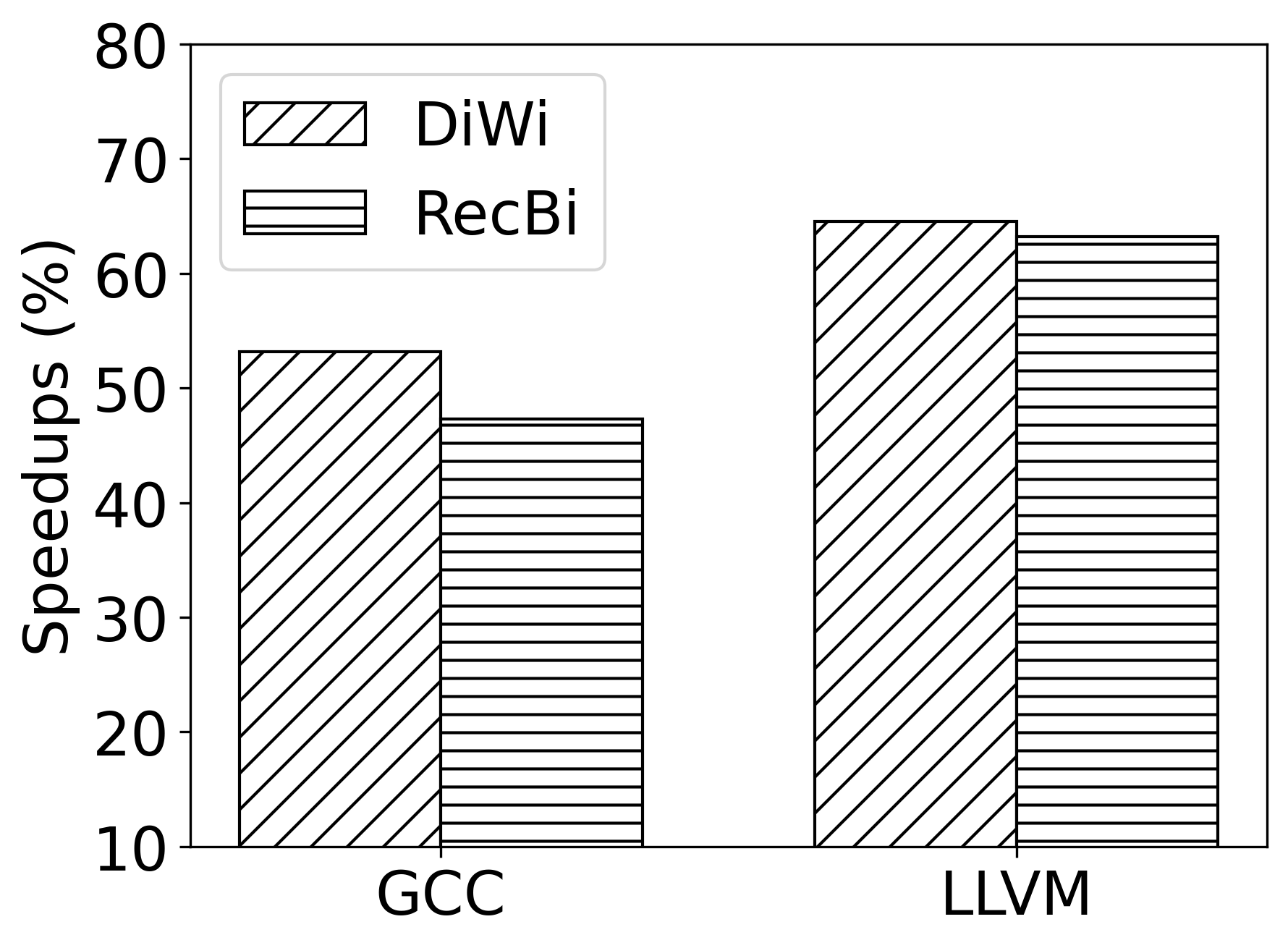}
\vspace{-1em}
\label{fig:speedups}
\end{minipage}	
}
\vspace{-1em}
\caption{Performance comparison over DiWi, RecBi, and \ourSol (under Setting-2 in RQ1)}
\label{fig:performance}
\end{figure}

For the efficiency side, Fig. \ref{fig:performance} presents the detailed performance results. The {\it x-axis} in box Fig. \ref{fig:time} show different approaches in GCC and LLVM, and {\it y-axis} represents the time spent by isolating each bug. For the figures, we can observe \ourSol takes less time when isolating most of the bugs. We can see the time taken on LLVM is generally longer than on GCC. This is justifiable as the number of files of LLVM is larger than GCC, and the more files the more time spent on calculating the coverage. We also calculate the speedups achieved by \ourSol. Specifically, we follow the Equation below to calculate the speedups:
\begin{equation}
\frac{T_{baseline} - T_{\ourSol}}{T_{baseline}} \times 100
\end{equation}
where $T_{baseline}$ represents the time spent by the baselines (i.e., DiWi and RecBi) for generating 10 passing test programs on average, and $T_{\ourSol}$ describes the time our proposed \ourSol spent generating the same number of passing test programs on average. Fig. \ref{fig:speedups} shows the results. We can see \ourSol is able to achieve 53.19\% and 64.54\% as well as 47.31\% and 63.19\% performance gains than DiWi and RecBi, for GCC and LLVM, respectively.

It is interesting to note that the overall results of \ourSol are better in Setting-1 compared with Setting-2. This is reasonable as \ourSol could generate more passing test programs in one hour, improving the results. Therefore, we suggest developers run a long time of \ourSol to get better isolation results.

\begin{table}[t]
	\centering
    \scriptsize
	\caption{Statistical Test Results under Setting-1 and Setting-2 \\ (\ourSol vs state-of-the-art approaches)}
        \vspace{-1em}
	\begin{tabular}{ccccccc}
		\toprule
            \multirow{2}*{\textbf{Approaches}} & 
		\multirow{2}*{\textbf{Metrics}} & 
            \multicolumn{2}{c}{\textbf{Setting-1}} &
		\multicolumn{2}{c}{\textbf{Setting-2}}  \\
		\cmidrule{3-6}
		&  &  \textit{p-value}& \textit{$\hat{A}_{12}$} & \textit{p-value}& \textit{$\hat{A}_{12}$}\\
		\midrule
            \multirow{6}{*}{ DiWi \cite{diwi} }
		& Top-1  & <0.0001 & 0.98 & <0.0001 & 1.00 \\
		& Top-5  & <0.0001 & 0.96 & <0.0001 & 0.98 \\
		  & Top-10 & <0.0001 & 0.97 & <0.0001 & 0.95 \\
            & Top-20 & <0.0001 & 0.96 & <0.0001 & 0.99 \\
            & MFR & <0.0001 & 1.00 & <0.0001 & 1.00 \\
            & MAR & <0.0001 & 1.00 & <0.0001 & 1.00 \\
            \midrule
             \multirow{6}{*}{ RecBi \cite{recbi} }
            &Top-1  & 0.0034 & 0.77 & 0.0004 & 0.82 \\
		&Top-5  & 0.0200 & 0.71 & 0.0355 & 0.69 \\
		  &Top-10 & 0.0015 & 0.79 & 0.0470 & 0.68 \\
            &Top-20 & 0.0001 & 0.85 & <0.0001 & 0.89 \\
            & MFR & <0.0001 & 1.00 & <0.0001 & 1.00 \\
            & MAR & <0.0001 & 1.00 & <0.0001 & 1.00 \\
		\bottomrule
	\end{tabular}
	\vspace{-1em}
	\label{tab:rq1-pvalue}
\end{table}

\begin{table*}
\renewcommand{\arraystretch}{1.0}
\scriptsize
\centering
\setlength{\tabcolsep}{4.5pt}
\caption{Compiler bug isolation effectiveness comparison with four variants of \ourSol}
\vspace{-1em}
\begin{threeparttable}
\begin{tabular}{c|c|cc|c c|c c|c c|c c|c c}
\toprule 
\multirow{2}{*}{\textbf{Subject}} &\multirow{2}{*}{\textbf{Approach}} & \textbf{Num.} &\textbf{$\Uparrow_{Top-1}$} & \textbf{Num.}& \textbf{$\Uparrow_{Top-5}$} & \textbf{Num.}& \textbf{$\Uparrow_{Top-10}$} & \textbf{Num.} & \textbf{$\Uparrow_{Top-20}$} &\multirow{2}{*}{\textbf{MFR}} & \textbf{$\Uparrow_{MFR}$} &  \multirow{2}{*}{\textbf{MAR}} & \textbf{$\Uparrow_{MAR}$} \\
& &  \textbf{Top-1} & (\%) & \textbf{Top-5 }&  (\%) & \textbf{Top-10 }&  (\%) & \textbf{Top-20} & (\%) &   & (\%) &  & (\%)\\
\midrule 
\multirow{5}{*}{ \textbf{GCC} } 
 &	\existPrompt	&	5.60	&	66.07	&	22.40	&	11.16	&	32.80	&	10.98	&	40.90	&	9.54	&	18.74	&	15.47	&	19.23	&	14.94	\\
&	\specficPrompt	&	7.10	&	30.99	&	23.80	&	4.62	&	34.30	&	6.12	&	40.10	&	11.72	&	18.73	&	15.43	&	19.20	&	14.83	\\
&	\randSel	&	6.40	&	45.31	&	21.80	&	14.22	&	31.20	&	16.67	&	41.60	&	7.69	&	18.69	&	15.25	&	19.24	&	15.00	\\
&	\selNoVal	&	7.70	&	20.78	&	22.40	&	11.16	&	33.20	&	9.64	&	39.70	&	12.85	&	18.92	&	16.26	&	19.27	&	15.15	\\
&	\ourSol	&	9.30	&	-	&	24.90	&	-	&	36.40	&	-	&	44.80	&	-	&	15.84	&	-	&	16.35	&	-	\\
\midrule \multirow{5}{*}{ \textbf{LLVM} } 
  &	\existPrompt	&	6.60	&	13.64	&	20.70	&	9.66	&	30.60	&	12.42	&	41.70	&	15.11	&	17.93	&	14.32	&	18.02	&	14.26	\\
&	\specficPrompt	&	6.20	&	20.97	&	20.70	&	9.66	&	31.90	&	7.84	&	41.50	&	15.66	&	17.50	&	12.24	&	17.63	&	12.34	\\
&	\randSel	&	6.20	&	20.97	&	21.50	&	5.58	&	33.40	&	2.99	&	42.90	&	11.89	&	16.20	&	5.17	&	16.27	&	5.03	\\
&	\selNoVal	&	5.90	&	27.12	&	20.80	&	9.13	&	33.10	&	3.93	&	43.90	&	9.34	&	17.97	&	14.53	&	18.09	&	14.60	\\
&	\ourSol	&	7.50	&	-	&	22.70	&	-	&	34.40	&	-	&	48.00	&	-	&	15.36	&	-	&	15.45	&	-	\\
\midrule \multirow{5}{*}{ \textbf{ALL}} 
   &	\existPrompt	&	12.20	&	37.70	&	43.10	&	10.44	&	63.40	&	11.67	&	82.60	&	12.35	&	18.34	&	14.91	&	18.62	&	14.61	\\
&	\specficPrompt	&	13.30	&	26.32	&	44.50	&	6.97	&	66.20	&	6.95	&	81.60	&	13.73	&	18.12	&	13.89	&	18.41	&	13.64	\\
&	\randSel	&	12.60	&	33.33	&	43.30	&	9.93	&	64.60	&	9.60	&	84.50	&	9.82	&	17.45	&	10.57	&	17.76	&	10.43	\\
&	\selNoVal	&	13.60	&	23.53	&	43.20	&	10.19	&	66.30	&	6.79	&	83.60	&	11.00	&	18.45	&	15.42	&	18.68	&	14.88	\\
&	\ourSol	&	16.80	&	-	&	47.60	&	-	&	70.80	&	-	&	92.80	&	-	&	15.60	&	-	&	15.90	&	-	\\
 \bottomrule
\end{tabular}
 \begin{tablenotes}
        \footnotesize
        \item Note: Columns “$\Uparrow*$” present the improvement rates (\%) of \ourSol over the compared approaches in terms of various metrics.
      \end{tablenotes}
  \end{threeparttable}
  \label{tab:rq2}
\end{table*}

{\bf Statistical Test Analysis of Results}.
After collecting experimental results of 10 runs of each approach, we follow existing works \cite{klees2018evaluating,jiang2021ctos} to conduct the Mann-Whitney U-test with a level of significance of 0.05 on the total bugs
between \ourSol and the state-of-the-art approaches according to the suggestions by Arcuri and Briand \cite{arcuri2011practical,vargha2000critique}. To further reduce the threats from randomness, we also calculate the effect size of the differences between \ourSol and the baselines using the Vargha and Delaneys $\hat{A}_{12}$ statistics. In our context, given a performance measure M (e.g., Top-1), the $\hat{A}_{12}$ statistics measure the probability that running approach A (e.g., \ourSol) yields higher M values than running another approach B (e.g., DiWi). If the two approaches are equivalent, then $\hat{A}_{12}$ = 0.5. For example, $\hat{A}_{12}$ = 0.9 entails we would obtain higher results 90\% of the time with approach A than approach B. We use the function ``wilcox.test(X,Y)'' written by  {\it R} language\footnote{\url{https://www.rdocumentation.org/packages/stats/versions/3.6.2/topics/wilcox.test}} to perform a Mann-Whitney U-test to get the {\it p-value}. For the effective size, we use the formula  $\hat{A}_{12} = (R_1 / m - (m + 1) / 2)  / n$ to calculate it,
where $R_1$ is the rank sum of the first data group we are comparing. The rank sum is a basic component in the Mann-Whitney U-test, and most statistical tools provide it. We use the function ``sum(rank(c(X,Y))[seq\_along(X)])'' written by {\it R} language to calculate the $R_1$ value, where X and Y are the data sets with the observations (i.e., the metrics over GCC and LLVM bugs) of the two compared randomized approaches.

The overall statistical results in both two experiment settings are shown in Table \ref{tab:rq1-pvalue}. The {\it p-value} less than 0.05 in the table shows that \ourSol performs significantly better than the state-of-the-art approaches, i.e., Diwi \cite{diwi} and RecBi \cite{recbi}. Furthermore, we can observe that all the effect sizes are greater than 0.71 for Setting-1 and 0.68 for Setting-2, which indicates that \ourSol has a higher probability of obtaining better results than the two state-of-the-art approaches.

\smallskip
\noindent
\textbf{Summary for RQ1.}
The extensive evaluation demonstrates that \ourSol significantly outperforms two state-of-the-art approaches in two different settings: \ourSol is able to effectively and efficiently generate witness test programs for compiler bug isolation than DiWi and RecBi.

\subsection{Answers to RQ2} \label{sec:answer2}

\subsubsection{Experimental Settings}

We use the following variants of \ourSol aiming to differentiate the effects of main components in \ourSol:
\begin{itemize}[leftmargin=1em,nosep]
    \item \existPrompt uses a simple prompt without data flow and control flow analysis. That is, \existPrompt only keeps the \colorbox{light-gray}{\makebox(66, 5){\textless mutation rule \textgreater}} and does not rely on the \colorbox{light-green}{\makebox(48, 5){\textless variables \textgreater}} and \colorbox{light-blue}{\makebox(42, 5){\textless location \textgreater}} information to fill in the designed pattern.
    \item \specficPrompt utilizes a specific prompt by giving the ``the most complex data and control flow''. That means, \specficPrompt replaces the \colorbox{light-green}{\makebox(48, 5){\textless variables \textgreater}} with ``the most complex variables'' and \colorbox{light-blue}{\makebox(42, 5){\textless location \textgreater}} with ``in the most complex statements'', which depends on the program understanding capabilities of LLMs to fill in the pattern.
    \item \randSel randomly selects a prompt without the memorized prompt selection, meaning \randSel performs a random selection of the prompt during the prompt selection process.
    \item \selNoVal removes the test program validation component in the prompt selection process in \ourSol: \selNoVal does not care about test program validity for compiler bug isolation.
\end{itemize}

Among these variants, \existPrompt and \specficPrompt aim to investigate whether the new program complexity-guided prompt production pattern is effective, while \randSel and \selNoVal target to understand whether the memorized prompt selection and test program validation could contribute to \ourSol, respectively.

\smallskip
\noindent
\textbf{Comparison Strategies}.
We run the four variants with the same strategy as Setting-1 in RQ1.
We then compare the average of 10 runs of Top-N, MFR, and MAR metrics for each approach to articulate the contribution of those components. 
Note that we run \ourSol, two state-of-the-art approaches, and four comparative approaches, 10 runs for each bug within 120 compiler bugs under two testing scenarios. Therefore, it takes more than 50 days to run the experiments under RQ1 and RQ2, indicating our experiment is extensive and sufficient.

\subsubsection{Experimental Results}

The comparison results between \ourSol and the comparative variant approaches are presented in Table \ref{tab:rq2}, where each column and row has the same meaning as in Table \ref{tab:rq1-1}. 

\smallskip
\noindent
{\bf Contribution of Prompt Production}.
As shown in Table \ref{tab:rq2}, \ourSol is better than both \existPrompt (i.e., the approach that uses a simple prompt without data flow and control flow analysis) and \specficPrompt (i.e., the approach that utilizes specific prompt by giving the ``the most complex data and control flow'') in terms of all the metrics over all the 120 bugs in GCC and LLVM.
Notably, \ourSol is able to isolate more 37.70\% and 10.44\% bugs within the Top-1 and Top-5 files and also improves 14.91\% and 14.61\% than \existPrompt in terms of MFR and MAR.
Compared with both \existPrompt and  \specficPrompt, \ourSol shows better results on Top-N metrics. In addition, \ourSol also holds a smaller value of MFR and MAR, indicating \ourSol has a better compiler bug isolation capability.

The above results indicate that the program complexity metrics help produce precise prompts, and removing data flow and control flow analysis or replacing them with explicit descriptions to LLMs are not effective. This is reasonable, as LLMs have shown to be limited at semantic understanding \cite{ma2023scope}. Therefore, with a limited prompt description, LLMs may randomly select interesting variables and locations when mutating the failing test program, making it difficult to flip the execution from failing to passing.

\smallskip
\noindent
{\bf Contribution of Prompt Selection}.
We run \randSel (i.e., an approach that randomly selects a prompt) and \ourSol to investigate the contribution of the memorized prompt selection component. Table \ref{tab:rq2} presents the overall results. The results indicate that \ourSol outperforms \randSel as well. Specifically, \ourSol demonstrates better performance compared to \randSel, with substantial improvement of 33.33\%, 9.93\%, 9.60\%, 9.82\%, 10.57\%, and 10.43\% across Top-1, Top-5, Top-10, Top-20, MFR, and MAR metrics, respectively.
These results underscore the superiority of our memorized prompt selection based on reinforcement learning over the random strategy. 

\begin{figure}[t]
\centering
\includegraphics[width=0.55\linewidth]{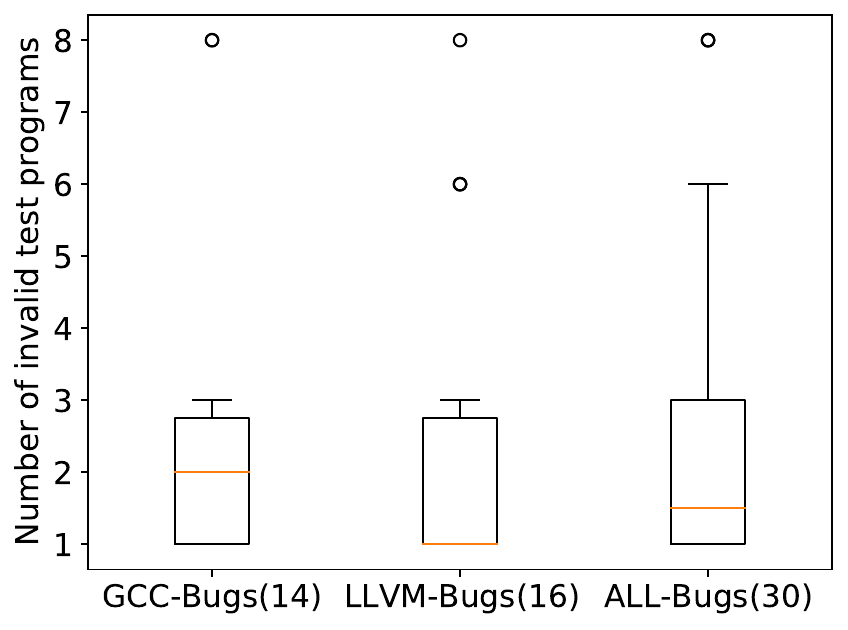}
\vspace{-1em}
\caption{The number of invalid programs generated by \selNoVal.}
\label{fig:ub-bar}
\end{figure}

\begin{figure}[t]
\centering
\begin{minipage}{.22\textwidth}
\centering
\begin{lstlisting}[caption={{\scriptsize LLVM Bug \href{https://bugs.llvm.org/show_bug.cgi?id=16041}{\#18000}: {\tt invalid shift} in Line 11 caused by the mutation in the same line}},basicstyle=\scriptsize,escapechar=@]
int a, b, c, d; char e;
void foo () { a = 0; }

int main (){
  unsigned char f;
  for (; b < 1; b++){  
    for (e = 1; e >= 0; e--){
      d = 0;
      if (a){ break; }
      f = 179 * e;
      @\colorbox{cyan}{\makebox(30,2){c = f << (-1);}}@ // c=f<<1;
      foo ();
    }
  }
  printf ("%d\n", c);
  return 0;
}
\end{lstlisting}
\end{minipage}%
\hspace{1.5em}
\begin{minipage}{.22\textwidth}
\centering
\begin{lstlisting}[caption={{\scriptsize GCC Bug \href{https://gcc.gnu.org/bugzilla/show_bug.cgi?id=64682}{\#64682}: {\tt invalid memory access} in Line 12 caused by the mutation in Line 2}},basicstyle=\scriptsize,escapechar=@]
int a;
short int b = 1; // int b = 1;

int main () {
  int i;
  for (i = 0; i < 56; i++) {
    for (; a; a--) { 
      ;
    }
  }
  int *c = &b;
  @\colorbox{cyan}{\makebox(15,2){if (*c)\{}}@
    *c=1 % (unsigned int)*c|5;
  }
  printf ("%d\n", b);
  return 0;
}
\end{lstlisting}
\end{minipage}
\vspace{-1em}
\caption{Side effects of the test programs with undefined behavior (the commented code in green is the original code in the failing test program, and the code with the \colorbox{cyan}{\makebox(30,2){cyan box}} contains undefined behavior)}
\label{fig:ub}
\end{figure}

\begin{figure}[t]
\centering
\subfigure[GCC benchmarks]{
\begin{minipage}[t]{0.97\linewidth}
\centering
\includegraphics[width=7cm]{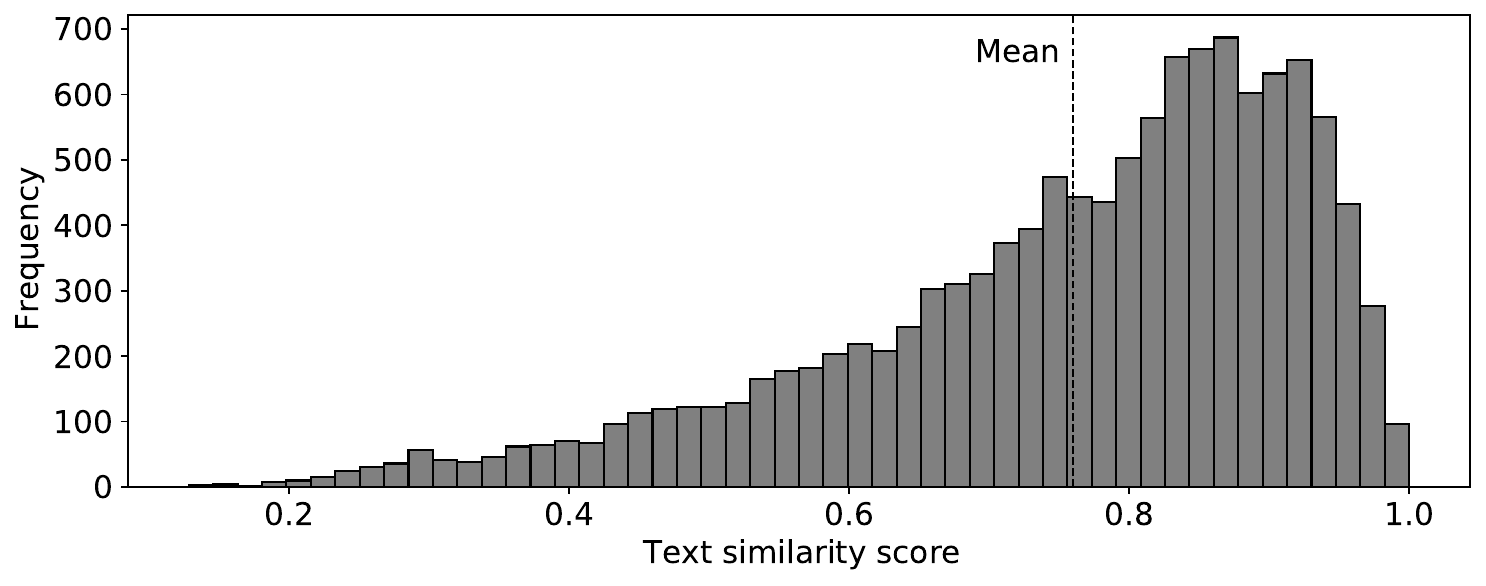}
\label{fig:var-scatter}
\end{minipage}
}
\subfigure[LLVM benchmarks]{	
\begin{minipage}[t]{0.97\linewidth}
\centering
\includegraphics[width=7cm]{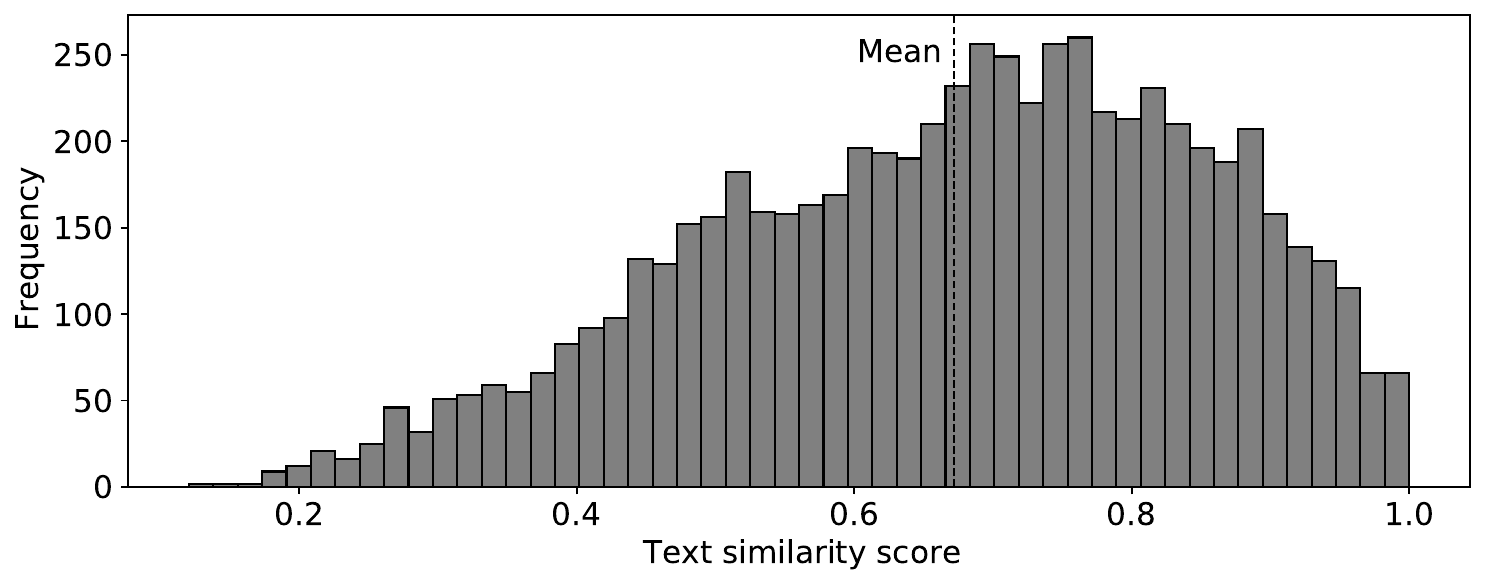}
\label{fig:var-box}
\end{minipage}	
}
\vspace{-1em}
\caption{Distribution of text similarity score among the test programs generated by \ourSol and \selNoVal}
\label{fig:sim-dist}
\end{figure}

\smallskip
\noindent
{\bf Contribution of Test Program Validation}.
Compared with the approaches shown in rows 4 and 5 in Table \ref{tab:rq2}, \ourSol outperforms \selNoVal (a variant approach that removes the test validation component) for all the metrics for both GCC and LLVM. 
For the overall benchmarks, \ourSol could isolate 23.53\% and 10.19\% more bugs with Top-1 and Top-5 files than \selNoVal, respectively.
In terms of MFR and MAR, \ourSol also exhibited significant improvement of 15.42\% and 14.88\% over \selNoVal. 
These results provide compelling evidence that incorporating the test program validation component enhances the effectiveness of compiler bug isolation, thereby confirming its valuable contribution in \ourSol.

\ourSol outperforms \selNoVal mainly because \selNoVal suffers from two severe issues that decrease its performance of bug isolation due to the existence of undefined behaviors, while LLVM4CBI could effectively reduce the threats by filtering those test programs.

The first issue comes from the unreliable test oracle. Typically, \ourSol treats the generated test program producing consistent results as a passing program, which serves as the test oracle in this study. However, since test programs generated by \selNoVal may contain undefined behaviors, the results they produced may not be reliable as the programs with undefined behaviors can have unexpected behaviors. If that happens, although the results show the program is a passing test program, it may be reliable to conclude that such a test program cannot trigger the bug, which disturbs the calculation of suspicious scores of files.

The second issue lies in the interferences between the compiler implementation code that handles the undefined behaviors and induces the compiler bugs. To be specific, it is well-known that C compiler developers hold the assumption that the programmer should not submit code that has undefined behaviors, and they implement optimization code under that assumption \cite{wang2012undefined}. For programmers who accidentally write code that has undefined behaviors, the compiler may remove the code (e.g., removing an access control check) or rewrite the code in a way that the programmer did not anticipate, which can result in an unexpected program behavior being compiled. Therefore, it is reasonable that the implementation of handling undefined behaviors may have intersections with the implementation that leads to compiler bugs, which may significantly affect the effectiveness of bug isolation. For example, if the compilers use the same utils functions to handle the undefined behaviors and compile the failing test program, the suspicious score of those utils functions could be low as such code is covered in passing test programs but not failing test programs. That means the real faulty functions will be silently hidden, thus reducing the chance of locating the exact functions that lead to the compiler bug. It is worth noting that how compilers handle undefined behaviors is a controversial topic in the literature and reliable identification of undefined behaviors in test programs is challenging \cite{wang2013towards,csmith,diwi}. We will continue to investigate this issue in future work.

To understand how many more valid test programs can be generated by \ourSol compared with \selNoVal, we run another set of experiments to confirm the contribution of filtering out invalid test programs. To conduct a fair comparison, we run \ourSol and \selNoVal in Setting-2, i.e., both approaches terminate when 10 passing test programs are generated. The experimental results are shown in the box plot in Fig. \ref{fig:ub-bar}. The x-axis represents the number of bugs in the benchmarks that contain invalid test programs of GCC bugs, LLVM bugs, and ALL(GCC+LLVM) bugs, where the number in the following brackets represents the number of bugs in the benchmarks that generate invalid test programs. We can observe that there are 30 (14 for GCC and 16 for LLVM) bugs in total that \selNoVal generates invalid test programs for them. The y-axis represents the distribution of the number of invalid test programs among those benchmarks, and the results show the average number of invalid test programs for GCC, LLVM, and ALL are 2.57, 2.50, and 2.53, respectively.  This means \ourSol is able to generate 2.5 more valid test programs than \selNoVal in the same experimental setting. Compared with all valid test programs generated by \ourSol, those invalid test programs decrease the bug isolation capabilities of \selNoVal.

To further understand why filtering the semantically invalid test programs is important, we showcase two examples with an undefined behavior in Fig. \ref{fig:ub} and indicate their side effects for compiler bug isolation.
The first example shown in Fig. \ref{fig:ub}(1) includes a kind of undefined behavior \texttt{shift}, which is generated by the mutation rule replacing a constant value (from ``1'' to ``-1'') in Line 12. RecBi ranked this bug at 23$^{th}$ due to the interference of the undefined behavior. The second example exposes a kind of undefined behavior \texttt{memem\_access}, which is produced by the mutation rule inserting a modifier of a variable (from \texttt{int} to \texttt{short int}) in Line 2. Because the size of \texttt{short int} (2 bytes in 64-bit system) is shorter than \texttt{int} (4 bytes in 64-bit system), there is an out-of-bound read from the variable \texttt{b}, which makes RecBi rank this bug at 28$^{th}$. Benefiting from the test program validation process, \ourSol filters those examples before ranking the suspicious files and finally ranks the LLVM bug and GCC bug at the position of 9$^{th}$ and 5$^{th}$, respectively. This is reasonable, as demonstrated in previous studies \cite{csmith,tu2022detecting,towards-understanding}, compilers can have unexpected consequences if the program is semantically invalid.

\smallskip
\noindent
{\bf Overlapping Analysis}.
To show whether the valid programs generated by \ourSol and \selNoVal are overlapping, we compare text and code coverage similarity between those valid programs. The text similarity shows whether two programs are similar in terms of content, and the code coverage similarity indicates whether two programs are similar in covering the compiler source code, as the programs with different contents may share the same code coverage in compilers. To calculate the text similarity, we use a well-established technique, i.e., TF-IDF (Term Frequency-Inverse Document Frequency), to compute the similarity score\footnote{We used the code from \url{https://github.com/bysiber/text_similarity_tfidf} to calculate the similarity score.} between two source code files from \ourSol and \selNoVal, where the similarity score is determined using cosine similarity \cite{manning2008introduction}. Note that the cosine similarity score ranges from ``0'' to ``1''. ``1'' means the two comparative source files are identical, while ``0'' indicates no similarity.

Fig. \ref{fig:sim-dist} shows the overall text similarity score among GCC and LLVM bugs, where the x-axis represents the score, and the y-axis denotes the frequency of the scores. To be specific, we compare every test program generated by \ourSol and \selNoVal element-wise, leading to 11990 and 6525 comparisons in total for GCC and LLVM bugs. As a result, the mean value of the score is 0.76 for GCC and 0.67 for LLVM bugs. We can observe that the valid programs generated by \ourSol and \selNoVal are hardly overlapping. There are only 3 and 10 comparisons that hold the score 1, which indicates the overlapping can be negligible.
The comparison code coverage similarly shares the same results as text similarity: the results show there are only 15 and 53 comparisons of two comparative source files holding the same code coverage in compilers. We believe the overlapping can be negligible since only a small portion of the test programs share the same code coverage.
In short, based on the comparison results of text and code coverage similarity, the valid test programs generated by \ourSol and \selNoVal are not or with negligible overlapping.

In summary, the semantically invalid test programs can have side effects on the effectiveness of compiler bug isolation, and filtering them does help \ourSol improve the ranking results.

\begin{table}[t]
	\centering
    \scriptsize
	\caption{Statistical Test Results (\ourSol vs variant approaches)}
        \vspace{-1em}
	\begin{tabular}{cccccc}
		\toprule
		\textbf{Approaches}& \textbf{Metrics} & \textit{p-value}& \textit{$\hat{A}_{12}$} \\
		\midrule
            \multirow{6}{*}{ \existPrompt}
		& Top-1  & <0.0001 & 0.87  \\
		& Top-5  & 0.0006 & 0.81  \\
		  & Top-10 & <0.0001 & 0.91 \\
            & Top-20 & <0.0001 & 0.94  \\
            & MFR & <0.0001 & 0.92  \\
            & MAR & <0.0001 & 0.91 \\
            \midrule
            \multirow{6}{*}{\specficPrompt  }
            & Top-1  & 0.0014 & 0.79  \\
		& Top-5  & 0.0275 & 0.70  \\
		  & Top-10 & 0.0080 & 0.74 \\
            & Top-20 & <0.0001 & 0.94  \\
            & MFR & <0.0001 & 0.88  \\
            & MAR & <0.0001 & 0.89 \\
            \midrule
            \multirow{6}{*}{ \randSel }
            & Top-1  & <0.0001 & 0.88  \\
		& Top-5  & 0.0001 & 0.85  \\
		  & Top-10 & 0.0005 & 0.82 \\
            & Top-20 & <0.0001 & 0.88  \\
            & MFR & 0.0021 & 0.79  \\
            & MAR & 0.0071 & 0.75 \\
            \midrule
            \multirow{6}{*}{\selNoVal}
            & Top-1  & 0.0252 & 0.70  \\
		& Top-5  & 0.0015 & 0.79 \\
		  & Top-10 & 0.0066 & 0.75 \\
            & Top-20 & 0.0002 & 0.85  \\
            & MFR & <0.0001 & 0.91  \\
            & MAR & <0.0001 & 0.91 \\
		\bottomrule
	\end{tabular}
	\label{tab:rq2-pvalue}
\end{table}

\smallskip
\noindent
{\bf Statistical Test Analysis of Results}. We use the same methodology as RQ1 to calculate {\it p-value} and effective size $\hat{A}_{12}$ in this RQ. The overall statistical results are shown in Table \ref{tab:rq2-pvalue}. The {\it p-value} less than 0.05 in the table shows that \ourSol performs significantly better than the various variant approaches. Furthermore, we can observe that all the effect sizes are greater than 0.70, which indicates that \ourSol has a higher probability of obtaining better results than comparative variant approaches.

\smallskip
\noindent
\textbf{Summary for RQ2.}
All three components, including precise prompt production, memorized prompt selection, and lightweight test program validation, contribute to the effectiveness of \ourSol.

\subsection{Answers to RQ3} \label{sec:answer3}

\subsubsection{Experimental Settings}

\begin{table}[t!]
        \scriptsize
		\centering
		\caption{Evaluated open-source LLMs in \ourSol}
            \vspace{-1em}
		\begin{tabular}{cccc} 
			\toprule
			\textbf{Models}& \textbf{Size} & \textbf{Release-Date}  &\textbf{Popularity(GitHub)}  \\
			\midrule
			Alpaca \cite{alpaca}         & 7B & March 2023 & 25.0K star   \\
			Vicuna \cite{vicuna} & 7B  & March 2023  & 22.8K star     \\
                GPT4ALL \cite{gpt4all}         & 13B & March 2023  & 46K star   \\
			\bottomrule
			\label{tab:models}
		\end{tabular}
		\vspace{-1em}
\end{table}

Table \ref{tab:models} presents the evaluated LLMs in this study, with column \textbf{Sizes} reflecting the model sizes in billions of parameters, \textbf{Release-Date} showing when the LLM is released, and \textbf{Popularity} indicating the number of GitHub stars or users counted by 1st June 2023. We evaluate three of the most representative and popular LLMs.
We choose the above LLMs mainly because (1) they are very popular that the number of stars soared to 20k+ in a few months and (2) they are proven to be effective in the code generation task \cite{evalplus,llama,alpaca,gpt4all}.
Based on the selected LLMs, we design three new variant approaches, i.e., \alpLLMs, \vicLLMs, and \allLLMs, by replacing the LLMs in \ourSol to answer this RQ.

\smallskip
\noindent
\textbf{Setting Up Different LLMs}.
For the three LLMs, we download the GGML\footnote{\url{https://github.com/ggerganov/ggml} \label{refnote}} format of these models from HuggingFace website\footnote{\url{https://huggingface.co/}} and use the python binding \texttt{llama-cpp-python}\footnote{\url{https://github.com/abetlen/llama-cpp-python}} of \texttt{llama.cpp}\footnote{The goal of \texttt{llama.cpp} is to run LLMs using 4-bit integer quantization on a MacBook only with CPU.} to run these models on a machine that only supports CPU. Specifically, we use a web server supported by \texttt{llama-cpp-python}, which aims to act as a drop-in replacement for the OpenAI API. This feature allows us to use \texttt{llama.cpp} compatible models with any OpenAI-compatible client (language libraries, services, etc.).
Note that it would be easy and require little human effort to support any other interesting models in \ourSol. The only thing users need to do is to provide the {\tt GGML} format model, either directly download from the HuggingFace website or product using the detailed and actively maintained tutorials in \texttt{GGML}'s homepage\footref{refnote}, for \texttt{llama-cpp-python} web server.

\smallskip
\noindent
\textbf{Comparison Strategies}.
We run those variants within a one-hour limit and then compare the Top-N metrics for each approach. We do not repeatedly run those models mainly because we aim to demonstrate the extendability of replacing different LLMs in \ourSol.

\begin{table}[t]
\renewcommand{\arraystretch}{1.0}
\centering
\scriptsize
\setlength{\tabcolsep}{4pt}
\caption{Compiler bug isolation capability of different LLMs in \ourSol \\ ( with 1 hour time limit)}
\vspace{-1em}
\begin{threeparttable}
\begin{tabular}{c|c|c|c|c|c}
\toprule \textbf{Subject} &\textbf{ Approach} & \textbf{Top-1} & \textbf{Top-5} & \textbf{Top-10} & \textbf{Top-20} \\
\midrule 
\multirow{4}{*}{ \textbf{GCC} }
  & \alpLLMs &  1 & 10 & 16 & 22 \\
   & \vicLLMs & 5  & 15 & 19 & 27 \\
   & \allLLMs & 2 & 7 & 15 & 21 \\
   & \ourSol & 11 & 26 & 41 & 50 \\
\midrule \multirow{4}{*}{ \textbf{LLVM} } 
  & \alpLLMs & 1  & 9 & 18 & 23  \\
   & \vicLLMs & 1  & 9 & 19 & 32   \\
   & \allLLMs & 1 & 5 & 18 & 26  \\
   & \ourSol & 10 & 24 & 33  & 48 \\
\midrule \multirow{4}{*}{ \textbf{ALL} } 
  & \alpLLMs & 2  & 19 & 34 & 45 \\
   & \vicLLMs & 6  & 24 & 38  & 59 \\
   & \allLLMs & 3 & 12 & 33 & 47  \\
   & \ourSol & 21 & 48 & 74 & 98  \\
 \bottomrule
\end{tabular}
  \end{threeparttable}
  \label{tab:rq3}
\end{table}

\subsubsection{Experimental Results}
Table \ref{tab:rq3} shows that all the designed variant approaches, i.e., \alpLLMs, \vicLLMs, and \allLLMs, contribute to the compiler bug isolation task. The results indicate that different LLMs can have different capabilities for generating test programs for compiler bug isolation, and GPT-3.5 (used as the default model in \ourSol) achieves the best performance compared with other LLMs.
Several reasons made the performance of other LLMs not as good as GPT-3.5 used in \ourSol:
\begin{itemize}[leftmargin=1em,nosep]
\item Only giving suggestion responses. Many outputs from those models are suggestions telling users how to change the code rather than the actual code. For example, on isolating LLVM \#17388 with the prompt ``replacing a binary operator with another valid one on the variables in the list ['a', 'c']'', \alpLLMs responses, ``You can replace the binary operator `||` with the logical operator `\&\&`''. Such output did not directly contain the complete test programs, but that information is still useful for helping users manually write a good passing test program. 
\item Limited code generation capabilities. Although other LLMs are targeting a comparable performance compared with GPT-3.5 or GPT-4, they may still be limited by the scale of training data. For example, The model {\tt Alpaca} used {\tt text-davinci-003} model to generate 52K instruction data as training data, which may still be limited compared with the training data scale of GPT-3.5 (although OpenAI did not disclose the actual number).
\item Performance issues. Running LLMs typically requires a powerful GPU to get better performance. Since our experiments run only on CPUs without GPUs, the response time for other LLMs is larger. Based on our observation, the response time of getting an answer from those LLMs takes approximately 300 seconds, which significantly reduces the effectiveness of those approaches. In contrast, GPT-3.5 only invokes API for the code generation, and the response time is less than 10 seconds, which makes \ourSol have a better capability.
\end{itemize}

Although the results from other LLMs are not as good as GPT-3.5, we show that \ourSol is extensible. With the rapid development of more powerful LLMs, such as {\tt Falcon}\cite{falcon40b,penedo2023refinedweb}, we believe advanced techniques can mitigate the above limitations and further assist the compiler bug isolation task in future work.

\begin{table}[t]
\renewcommand{\arraystretch}{1.0}
\centering
\scriptsize
\setlength{\tabcolsep}{4pt}
\caption{Compiler bug isolation capability of different LLMs in \ourSol \\ (generating the same number of passing test programs)}
\vspace{-1em}
\begin{threeparttable}
\begin{tabular}{c|c|c|c|c|c}
\toprule \textbf{Subject} &\textbf{ Approach} & \textbf{Top-1} & \textbf{Top-5} & \textbf{Top-10} & \textbf{Top-20} \\
\midrule 
\multirow{5}{*}{ \textbf{GCC} }
  & DiWi \cite{diwi} & 3& 14& 25 & 34 \\
  & RecBi \cite{recbi} & 5& 21& 28&  38  \\
  & \alpLLMs & 6  & 22 &30  & 40 \\
   & \vicLLMs & 7  &  23& 31 & 41 \\
   & \allLLMs & 6 & 23 & 32 & 42 \\
\midrule \multirow{5}{*}{ \textbf{LLVM} } 
   & DiWi \cite{diwi} &1& 11& 25&  35 \\
   & RecBi \cite{recbi} &3   & 19 &  30& 37 \\
   & \alpLLMs &  4 & 20  & 31 & 39  \\
   & \vicLLMs &  6 & 21 & 33 &   40 \\
   & \allLLMs & 5 & 20 & 32 &  38 \\
\midrule \multirow{5}{*}{ \textbf{ALL} } 
    & DiWi \cite{diwi} &4& 25&50 & 69 \\
    & RecBi \cite{recbi} & 8 & 40 & 58 & 75 \\
  & \alpLLMs & 10  & 42 & 61  & 79 \\
   & \vicLLMs & 13  & 44  & 64  & 81 \\
   & \allLLMs & 11 & 43 & 64 & 80  \\
 \bottomrule
\end{tabular}
  \end{threeparttable}
  \label{tab:rq3-same-nu}
\end{table}

We would like to clarify that the ``extensible'' used in the paper is only meant from a software architecture perspective, and we define ``extensible'' as the capability and difficulty of replacing the default LLM (i.e., ChatGPT) of LLM4CBI with other LLMs. Regarding the results of bug isolation, we think it depends on the capabilities of those LLMs; certain limitations in LLMs prevent them from better compiler bug isolation, and we have discussed several reasons above (i.e., only giving suggestion responses, limited code generation capabilities, and performance issues) why some recent LLMs cannot yield good results. To further support the potential usefulness of replacing ChatGPT with other LLMs, we changed the running setting with termination with 1 hour to generate the same number (i.e., 5) of passing test programs, and the results are shown in Table 9. From the table, we can observe that using other LLMs can perform better than the two prior studies (DiWi and RecBi).

\smallskip
\noindent
\textbf{Summary for RQ3.}
\ourSol is extensible, meaning LLMs component (i.e., GPT-3.5) used in \ourSol can be easily replaced by other LLMs (e.g., Alpaca \cite{alpaca}, Vicuna \cite{vicuna}, and GPT4ALL \cite{gpt4all}) while still achieving reasonable results in comparison to related studies.

\section{Discussion} \label{sec:discussion}

In this section, we first introduce the practical implications of file-level compiler bug isolation, and then discuss the comparison results with GPT-4 as well as evaluate the impact of different {\it temperature} values used in \ourSol. Finally, we discuss some limitations of \ourSol.

\subsection{Practical Implications of File-level Bug Isolation} 
In this paper, we target file-level compiler bug isolation, and the file-level ranking results may not fully solve the compiler bug isolation problem, i.e., \ourSol currently is not capable of isolating faulty positions at line-level. However, due to the large number of optimizations (200+ fine-grained optimization passes for GCC and LLVM, respectively) and files (1,588 files and 3,507 files for GCC and LLVM, respectively) and sophisticated interplay between optimizations, pinpointing which file is buggy is still challenging. Furthermore, with the confirmation by the feedback from compiler developers in the previous work (e.g., DiWi \cite{diwi}), the majority (6 out of 7) communicated developers confirmed that their compiler debugging process starts from buggy file identification and this step is time-consuming, indicating the necessity of compiler bug isolation at the file level. Although some developers have an intuition about which part may be wrong, not all developers have such expertise, and their intuition may not be accurate. We are actively pursuing the design of a more fine-grained level (e.g., method) compiler bug isolation approach. As we do not know whether the training data of LLMs contains compiler bugs, we also plan to curate recent real-world compiler failure reports (so that it is less likely that they have been used as training data for LLMs) and evaluate the effectiveness of \ourSol for bug isolation on those failure cases with actual developers of GCC or LLVM.

We would like to emphasize the usability of bug isolation with file-level granularity. Compared with normal software systems, the implementation of compilers typically involves a vast number of source code files (a buggy GCC and LLVM compiler contains 1,758 and 3,265 files on average, respectively) and sophisticated intersection/interplay between different optimizations as well as complex implementation of single optimization component (a prior study shows only single peephole optimization component at least involve 10 files \cite{theodoridis2022finding}). The above facts indicate the increasing difficulty of locating compiler bugs in file granularity compared with normal software systems. For example, the failing test program of bug \#16041 involves 418 files in total, and every file could be suspicious of the bug. Furthermore, most compiler optimizations are not solely implemented, i.e., optimizations can interact in non-trivial ways, where the benefit of one optimization may affect others, meaning the dependencies of different files that implement optimizations could be exceedingly complex, indicating locating suspicious files for only one single optimization bug is intractable. Considering the large number of compiler optimizations, locating the file containing the root cause could be considerably difficult. As we demonstrated in the evaluation, \ourSol can isolate 16.80, 47.60, 70.80, and 92.80 compiler bugs (out of a total of 120 compiler bugs in GCC and LLVM) within the Top-1, Top-5, Top-10, and Top-20 files, respectively. For bug \#16041, among 418 covered files by failing test program, \ourSol ranks the faulty file at 6{\it th} position, which could help developers debug the bug efficiently.
Therefore, an effective file granularity bug isolation approach is still needed and useful for isolating bugs in complex software systems such as compilers.

\subsection{Comparison with GPT-4} 
In this study, we are unable to conduct large-scale experiments using GPT-4 in \ourSol due to its higher API cost\footnote{\url{https://openai.com/pricing}} compared to GPT-3.5. However, we did run two cases (LLVM bug 16040 and 25154) using a variant of \ourSol, called \gptFourLLMs, which incorporates GPT-4. Interestingly, the results demonstrated that \gptFourLLMs outperformed \ourSol, with both Top 20+ (i.e., Top-40 for bug 16040 and Top-58 for 25154) bugs being ranked within Top-5 (i.e., Top-3 for bug 16040 and Top-2 for 25154).
The improved performance of \gptFourLLMs can be attributed to two main factors. First, GPT-4 generates a smaller number of test programs with syntax errors, indicating a higher quality of output. Second, GPT-4 exhibits a better understanding of the prompt, leading to improved results \cite{evalplus,vicuna}.
As LLMs continue to advance rapidly, we anticipate the availability of more capable and user-friendly models that can be integrated into \ourSol, further enhancing the compiler bug isolation capabilities of \ourSol.

\subsection{Different {\it temperature} Settings in LLMs} 
\begin{figure}[t]
\centering
\subfigure[GCC]{
\begin{minipage}[t]{0.46\linewidth}
\centering
\includegraphics[width=4cm]{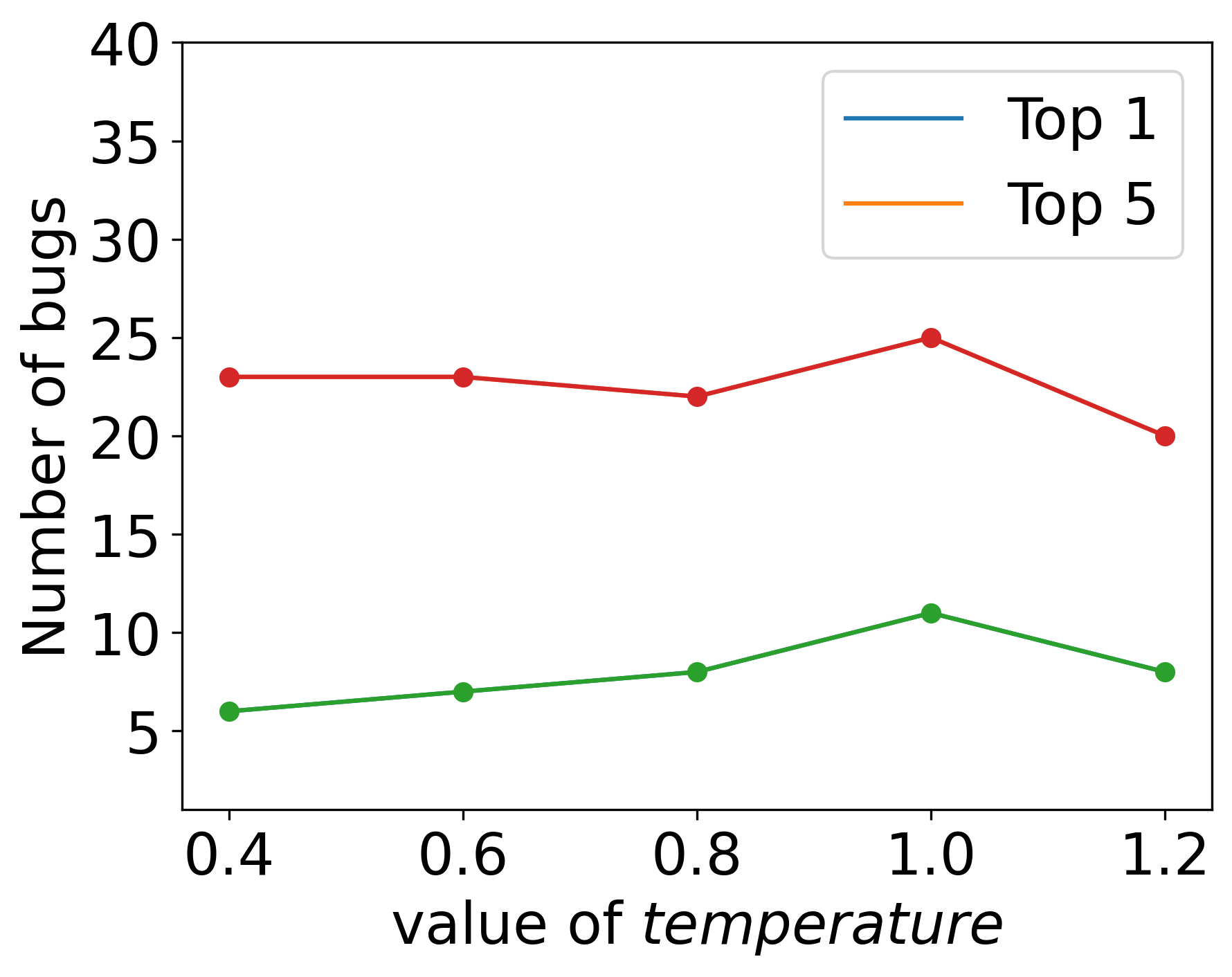}
\vspace{-1em}
\label{fig:temp1}
\end{minipage}
}
\subfigure[LLVM]{	
\begin{minipage}[t]{0.46\linewidth}
\centering
\includegraphics[width=4cm]{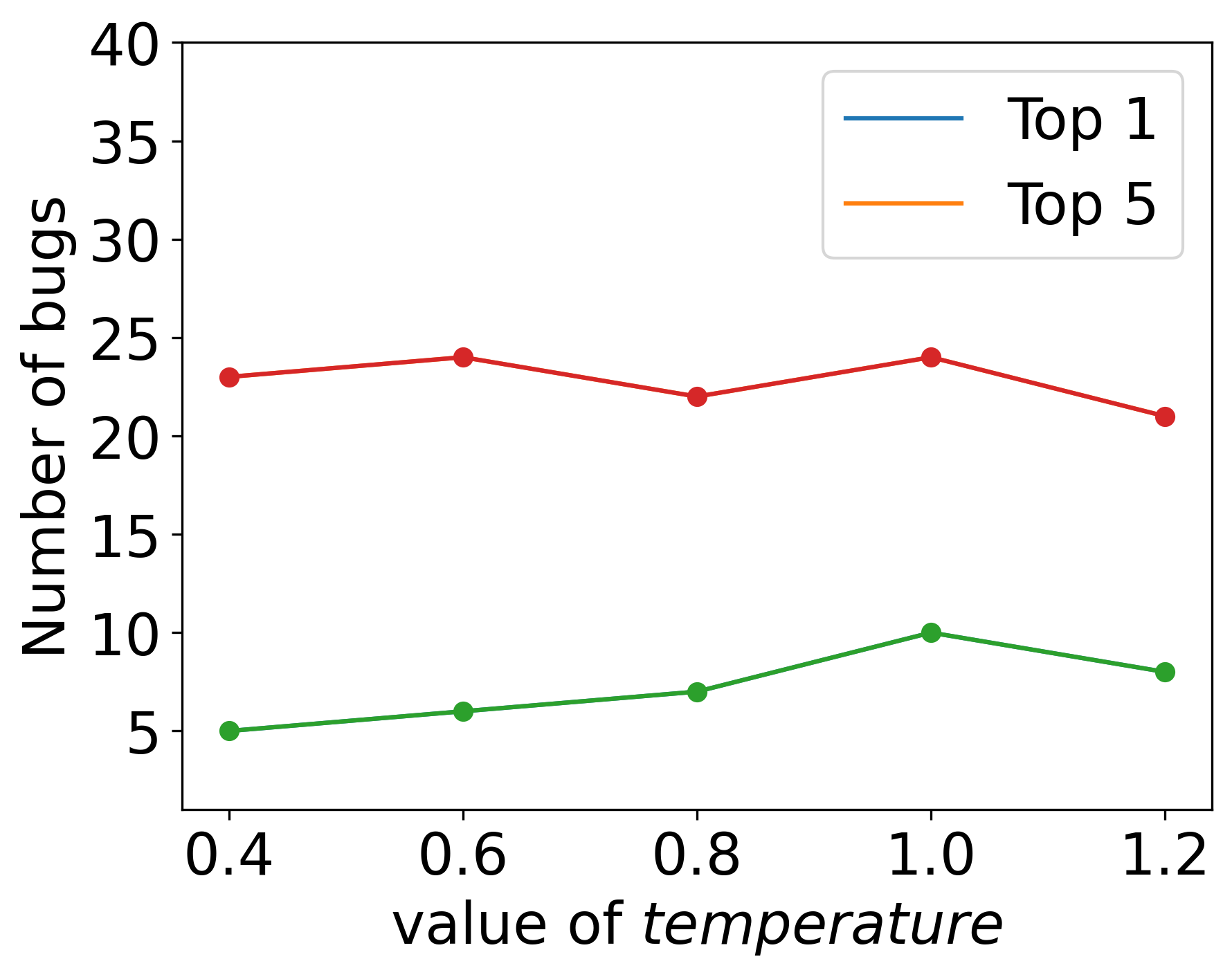}
\label{fig:temp2}
\end{minipage}	
}
\vspace{-0.5em}
\caption{The results of the impact of different {\it temperature} settings}
\label{fig:temp}
\end{figure}

The {\it temperature} parameter in LLMs, such as GPT-3.5, plays a crucial role in controlling the randomness of the generated output. A lower temperature value makes the output more deterministic, while a higher temperature value increases randomness. In the context of \ourSol, it is important to investigate the impact of different temperature values on the performance of test program generation for compiler bug isolation.
To this end, we conducted experiments using temperature values of 0.4, 0.6, 0.8, 1.0, and 1.2. The bug isolation capabilities of \ourSol were compared under these different temperature settings, and the results are shown in Fig. \ref{fig:temp}.
The results show that the default temperature value of 1.0 performs the best for \ourSol. This finding aligns with expectations since OpenAI has fine-tuned this parameter and set it as the default value, indicating that it yields optimal results.

\subsection{Impacts of selecting complex variables} 

\begin{figure}[t]
\centering
\subfigure[Scatter plot of the distribution]{
\begin{minipage}[t]{0.46\linewidth}
\centering
\includegraphics[width=4.4cm]{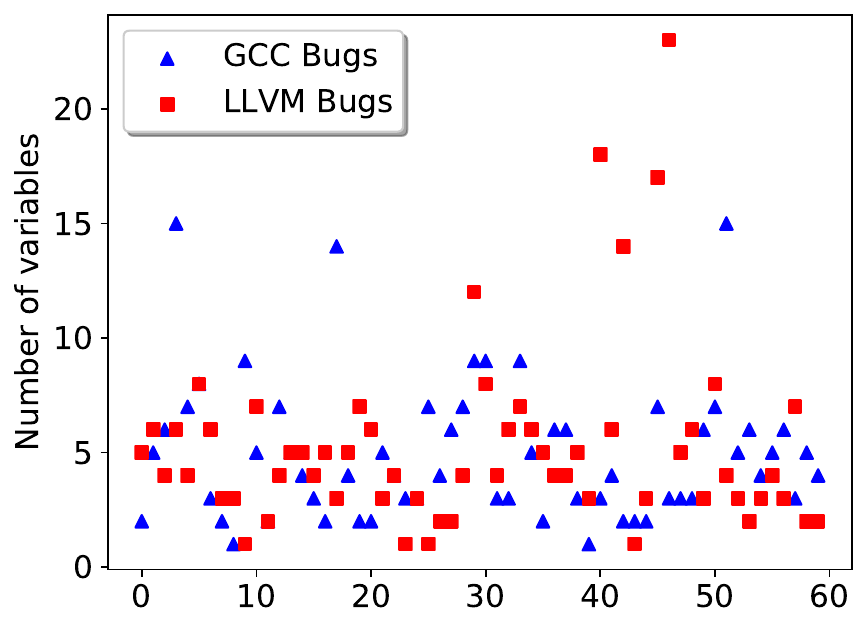}
\vspace{-1.5em}
\label{fig:var-scatter}
\end{minipage}
}
\subfigure[Box plot of the distribution]{	
\begin{minipage}[t]{0.46\linewidth}
\centering
\includegraphics[width=4.4cm]{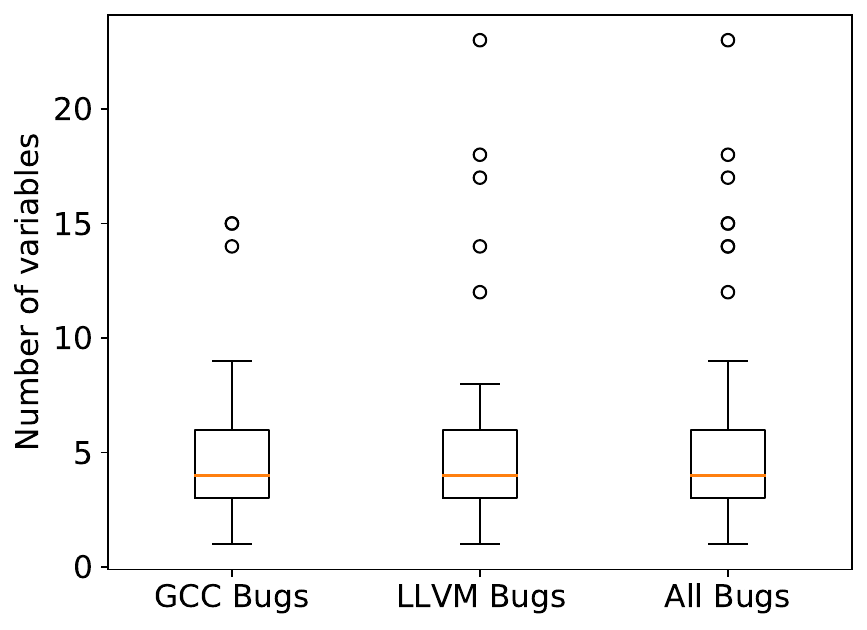}
\vspace{-1.5em}
\label{fig:var-box}
\end{minipage}	
}
\vspace{-0.5em}
\caption{Distribution of variable numbers among failing test programs}
\label{fig:var-dist}
\end{figure}

\begin{figure}[t]
\centering
\includegraphics[width=0.6\linewidth]{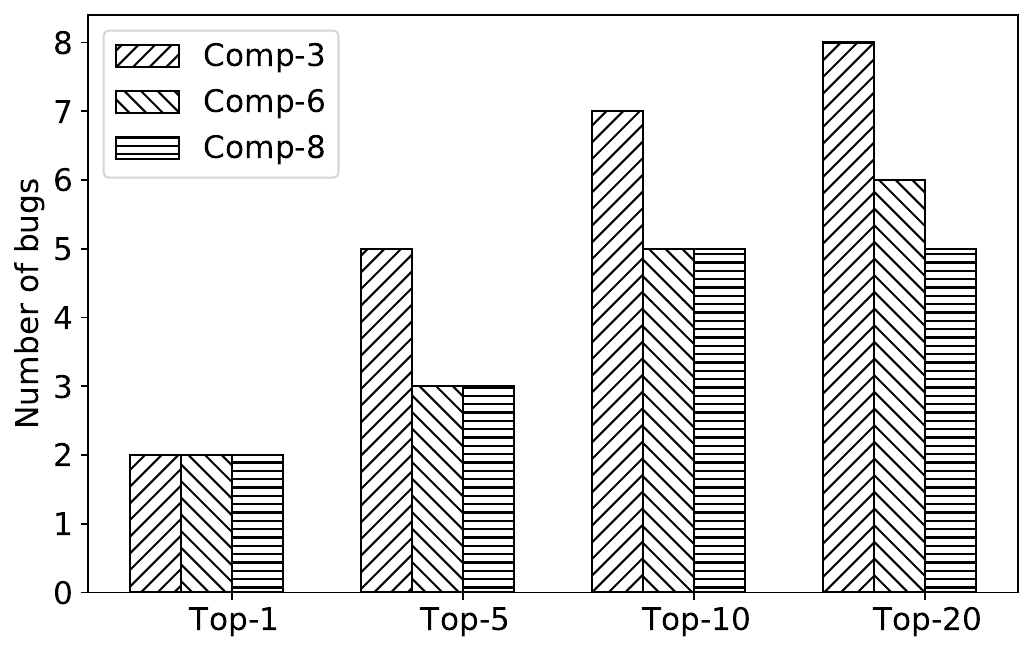}
\vspace{-1em}
\caption{Impact of selecting different numbers of complex variables}
\label{fig:var-bar}
\end{figure}

Since every failing test program contains a different number of variables to be selected, \ourSol selects a different number of complex variables. For the number of variables less or equal to 3, \ourSol selects the most complex variables ranking at the top one (denoted as Comp-1), and for the rest whose variables numbers are larger than 3, \ourSol selects the most complex variables ranked at top three (refers to Comp-3). Fig. \ref{fig:var-scatter} and Fig. \ref{fig:var-box} present the scatter and box plot of the distribution of variables over the 120 bugs. Both y-axis of Fig. \ref{fig:var-scatter} and Fig. \ref{fig:var-box} indicate the number of variables that can be selected, while the x-axis in Fig. \ref{fig:var-scatter} represents the order of all bugs in GCC and LLVM and the x-axis in Fig. \ref{fig:var-box} shows the GCC, LLVM, and All (GCC+LLVM) bugs. The average available and optional numbers of GCC and LLVM are 4.97 and 5.28, respectively. From the statistical point of view, \ourSol selects Comp-1 variables for 37.5\% bugs (24 for GCC bugs and 21 for LLVM bugs) and Comp-3 variables for the rest 62.5\% of the bugs, where ``Comp-{\it n}'' refers to the most complex variables ranked at the top {\it n} among the optional number of variables.
It is worth noting that in our experiments, LLMs generate valid passing programs for all 120 bugs under the above variable selection strategies, meaning that LLMs never fail at generating any valid witness program when selecting either Comp-1 or Comp-3 variables for different bugs.

To further understand how different numbers of selected variables affect the performance of \ourSol, we select 12 bugs whose optional numbers of variables are larger or equal to 9 and choose Comp-3/6/8 as the large set of complex variables to study the impacts. The evaluation results are shown in Fig. \ref{fig:var-bar}. We can observe that the setting of Comp-3's results is superior to the other two settings, indicating that selecting the most 3 complex variables is a better choice. The results are reasonable as changing the most complex variables is more likely to yield passing test programs, thus increasing the effectiveness of \ourSol, which is aligned with our initiation to select the most complex variables to fill in the pattern rather than randomly selecting variables like existing approaches.

\subsection{Limitations of \ourSol}
The limitations in \ourSol inherit the issues from SBFL techniques and LLMs. First, SBFL suffers from tie issues \cite{tie1,tie2}. A possible solution is to address this issue by incorporating the use of commit history information \cite{historical}.
Second, LLMs may generate syntactically invalid programs, reducing the effectiveness of \ourSol. However, based on the experimental results, the number of invalid test programs is acceptable and does not affect the results much. A more effective approach could leverage the program repair technique \cite{repairllm} to automatically repair the code generated by LLMs. We leave further investigation on the above two directions as our future work.

\section{Threats to Validity} \label{sec:threats}

This section discusses the {\it internal}, {\it external}, and {\it construct} threats of \ourSol.

The {\it internal} validity concerns stem from the implementation of \ourSol and comparative approaches (DiWi \cite{diwi} and RecBi\cite{recbi}). To mitigate this threat, we have extended the implementation provided by the previous studies \cite{diwi,recbi}. As for \ourSol, we have meticulously developed its implementation by leveraging well-established libraries, as explained in Section \ref{sec:implementation}, and have conducted thorough code checking of the code. For the tool used in test program validation, Frama-C \cite{framac} can not detect all the undefined behaviors in the programs as the identification of undefined behaviors in the programs is a challenging problem \cite{lee2017taming,wang2013towards}. We plan to leverage more advanced techniques to detect undefined behaviors in the test programs.

The major {\it external} validity of our study can be influenced by two key factors: the selection of compilers and the presence of bugs. To ensure the reliability of our findings, we have followed the established practices in prior studies on compiler bug isolation \cite{diwi,recbi} when selecting compilers. Following these approaches, we have employed two widely used open-source C compilers, GCC and LLVM, which are renowned for their popularity and extensive usage within the community.
Regarding the bugs used in our study, we chose a comprehensive set of 120 real compiler bugs, encompassing all known bugs from prior investigations in the field of compiler bug isolation. 
To further bolster the external validity of our research and address potential threats, we are committed to expanding our bug dataset by incorporating additional real compiler bugs. 

Another threat lies in the LLMs used in this study.
Specifically, the training data of LLMs (e.g., ChatGPT) has likely included the source code of GCC and LLVM, and maybe even some historical bug reports, bug-revealing test programs, and bug fixes. However, compared with the massive amounts of training data (more than 45 TB text \footnote{ \url{https://www.springboard.com/blog/data-science/machine-learning-gpt-3-open-ai/}}), the source code of compilers only takes a small part of among the training data. Therefore, if bugs used for training were also used in our testing, it is less likely to inflate the experimental results. Ideally, for testing, we should utilize new bugs that have for sure not been included as training data for ChatGPT; however, although there are such newly reported bugs in the wild, they are often not fixed yet and lack the ground truth for bug isolation, making it challenging to measure the isolation results. We also plan to curate recent real-world compiler failure reports (so that it is less likely that they have been used as training data for LLMs) and evaluate the effectiveness of \ourSol for bug isolation on those failure cases with actual developers of GCC or LLVM in future work.

The {\it construct} validity threats are subject to potential threats related to randomness, evaluation metrics, and parameter settings during the evaluation process. We address these concerns by (1) repeating experiments 10 times with a running time of more than 50 days in total and calculating average results to account for randomness, thereby reducing the impact of random variations; (2) employing widely-used bug localization metrics to assess the effectiveness of \ourSol to ensure the reliability and comparability of our evaluation results; (3) providing explicit parameter configurations and thoroughly investigating their impact in Section \ref{sec:discussion}. By doing so, we enhance transparency and enable a deeper understanding of the influence of different parameter settings.

\section{Related work} \label{sec:related-work}

This section surveys the most related works in this study, namely compiler debugging and LLMs for code generation.

\smallskip
\noindent
\textbf{Compiler Debugging.}
Our study is primarily related to two recent works: DiWi \cite{diwi} and RecBi \cite{recbi}. Both are general solutions for isolating all kinds of bugs in compilers. These works address the problem of isolating compiler bugs by transforming them into the generation of passing test programs. DiWi and RecBi achieve this by first utilizing mutation operators (DiWi focuses on local mutation operators while RecBi supports structural mutation operators) to generate a set of passing test programs similar to the failing test program, but without triggering the bug. Then, they employ spectrum-based bug localization techniques \cite{abreu2007accuracy,historical} to rank the buggy compiler files by comparing execution traces between the generated passing test programs and the failing test program.
LocSeq \cite{zhou2022locseq} focuses on isolating optimization bugs only in LLVM, while ODFL \cite{yang2022isolating} aims to isolate bugs only in GCC. We did not compare \ourSol with either LocSeq or ODFL mainly because they are not general enough and can only be applicable to specific middle-end compiler optimization bugs. In contrast, \ourSol is capable of isolating all kinds of bugs in compilers, including front-end and back-end bugs. 

In addition, Zeller \cite{zeller2002isolating} introduces a method to facilitate GCC debugging by calculating the cause-effect chain through a comparison of program states between a passing run and a failing run.
Holmes and Groce \cite{holmes2018causal,holmes2020using} propose a method to localize compiler bugs by comparing a set of compiler mutants. 
Regarding compiler debugging techniques for other programming languages, Chang et al.  \cite{chang2005type} present an approach for debugging the just-in-time compiler in a Java virtual machine. 
Lim et al. \cite{loc-jit-compiler1,loc-jit-compiler2} leverage dynamic analysis to locate bugs in JIT compilers.

Different from those works, we follow the existing strategy to generate a set of effective test programs for compiler bug isolation.
In contrast, we propose a new structural mutation strategy to mutate failing test programs effectively: we support complete control-flow statements mutation that includes both statement conditions and bodies. Besides, \ourSol only requires little human effort to accomplish the test program mutation process.

\smallskip
\noindent
\textbf{LLMs for Code Generation.}
The emergence of LLMs has sparked significant interest in their application to the code generation task.
Ling et al. \cite{ling2016latent} follow an {\it encoder-decoder} design and use a sequence-to-sequence LSTM model with attention and a copy mechanism to generate Java and Python programs. Iyer et al. \cite{iyer2018mapping} use a grammar-aware decoder to generate syntactically valid Java parse trees followed by Java codes using a two-step attention mechanism.
CodeBERT \cite{feng2020codebert} and GraphCodeBERT \cite{Guo2020GraphCodeBERTPC} inherit the design of BERT \cite{devlin2018bert} which are {\it encoder}-only models to generate codes. Moreover, CodeT5~\cite{wang2023codet5} and PLBART~\cite{DBLP:conf/naacl/AhmadCRC21} leverages encoder-decoder architecture for generating codes.
GLAsT \cite{harris2016glast} and VGen \cite{thakur2023benchmarking} apply {\it decoder}-only LLMs for generating Verilog RTL programs. Other recent LLMs, such as ChatGPT \cite{chatgpt}, Vicuna \cite{vicuna}, WizardLM \cite{wizardlm}, and StableLM \cite{stablelm}, all follow {\it decoder}-only architecture to generate various test programs in Python, Java, C/C++, and SQL.

In this study, we opt for the {\it decoder}-only models that follow the prompt-response dialog paradigm to generate the whole test program for the compiler bug isolation task. Different from existing approaches, we design a new pattern for precisely producing prompts in \ourSol. The program's data and control flow complexity are measured to fill in the designed pattern. Furthermore, a memorized selection and a test program validation strategy are proposed to select the proper prompt for effectively taming LLMs for test program mutation.
We also demonstrate \ourSol can be extensible for adopting different LLMs for compiler bug isolation.

\smallskip
\noindent
\textbf{Prompt Engineering for LLMs.}
Increasing number of studies \cite{liu2022fill,deng2023large,white2023prompt} adopt various strategies for effective prompt engineering to leverage the capabilities of LLMs. For example, Deng et al. \cite{deng2023large} propose TitanFuzz to directly leverage LLMs to generate input programs for fuzzing deep learning libraries. Given any target API, TitanFuzz first uses generative LLMs to generate a list of high-quality seed programs for fuzzing and then leverages the ability to infill LLMs to perform code infilling to generate new code that replaces the masked tokens, where the masked tokens analogous to the mutation operators in this study. During the mutation process, TitanFuz utilizes an evolutionary fuzzing algorithm to iteratively select new code snippets by replacing various masked tokens. Liu et. al \cite{liu2022fill} introduce QTypist to boost the performance of LLMs in the mobile testing scenario. QTypist extracts the context information for the text input and designs linguistic patterns to generate prompts for inputting into the LLMs. To further boost the performance of LLMs in mobile input scenarios, QTypist adopts a prompt-based data construction and tuning method, which automatically builds the prompts and answers for model tuning. 

Compared with existing prompt engineering in LLMs, two major differences between \ourSol and others. First, the formulation strategies of prompts are different. Since existing studies focus on generating diverse code/text for software testing, they typically share the same idea of random fuzzing. That means, they randomly select elements to fill in the prompt patterns. Even though TitanFuzz considers static analysis to build a fitness function for selecting better mutators, it does not take the control flow into account for more effective prompt engineering. Note that such strategies might be effective for software testing purposes as random testing has been shown good potential to detect program bugs. However, due to the lack of precise prompt instructions, it is not effective for the compiler bug isolation task as shown in our experiments. Second, existing works do not consider giving feedback to LLMs to refine the prompts, i.e., they only apply one-time execution of LLMs, and the learning capabilities of LLMs are not fully activated in those works. In contrast, \ourSol proposes a precise formulation of prompts by leveraging both data and control flow analysis. Furthermore, \ourSol applies reinforcement learning to refine the prompt and assist \ourSol continuously selecting better prompts.

\section{Conclusion and Future work} \label{sec:conclusion}

We have presented \ourSol, a new approach to tame LLMs for generating effective test programs for compiler bug isolation. In \ourSol,  three new components, i.e., precise prompt production, memorized prompt selection, and lightweight test program validation, are designed to tackle the two main challenges of formulating precise prompts and selecting specialized prompts.
Empirical evaluation using 120 real-world bugs from GCC and LLVM demonstrates the effectiveness of \ourSol over state-of-the-art approaches. Notably, \ourSol isolate 69.70\%/21.74\% and 24.44\%/8.92\% more bugs than DiWi and RecBi within Top-1/Top-5 ranked results. Furthermore, we demonstrate that \ourSol is extensible, highlighting its ease of extension to other LLMs for the compiler bug isolation task.

\smallskip
\noindent
\textbf{Future Work}.
In addition to the promising results presented in this paper, there are several exciting directions for further enhancing the capabilities of \ourSol. We are actively exploring the following directions:
\begin{itemize}[leftmargin=1em,nosep]
    \item Interactive bug isolation. Taking inspiration from interactive fault localization techniques that leverage user feedback \cite{gong2012interactive}, we may incorporate interactive elements into \ourSol. In detail, by treating LLMs (e.g., ChatGPT) as a user, we can enable them to learn continuously from the feedback received during the bug isolation process, thereby improving the isolation capability.
    \item Intelligent bug isolation. To empower LLMs with more detailed information, such as coverage data, we can enable them to make intelligent decisions regarding potentially buggy files. Specifically, we plan to guide LLMs to employ different evaluation strategies (e.g., different formulas used in SBFL) and fine-tune the parameters of LLMs to facilitate compiler bug isolation.
\end{itemize}

\ifCLASSOPTIONcaptionsoff
  \newpage
\fi

\section*{Acknowledgment}
The authors would like to thank all developers who participated in this work and the anonymous reviewers for their insightful comments. This work is supported in part by the National Natural Science Foundation of China (Grants No. 62032004 and No. 62302077) and China Postdoctoral Science Foundation (Grants No. 2023M730472). 

\bibliographystyle{IEEEtran}
\bibliography{references.bib}
\end{document}